\documentclass[aip,cha,amsmath,amssymb,reprint]{revtex4-2}
\usepackage{graphicx,xcolor}
\usepackage{bm,bbold}

\newcommand{\bigzero}{\mbox{\large 0}}
\newcommand{\rvline}{\hspace*{-\arraycolsep}\vline\hspace*{-\arraycolsep}}

\usepackage[utf8]{inputenc}
\usepackage[T1]{fontenc}
\usepackage{mathptmx}
\usepackage{etoolbox}

\usepackage{hyperref}
\usepackage{braket}
\usepackage{enumitem}

\makeatletter
\def\@email#1#2{%
 \endgroup
 \patchcmd{\titleblock@produce}
  {\frontmatter@RRAPformat}
  {\frontmatter@RRAPformat{\produce@RRAP{*#1\href{mailto:#2}{#2}}}\frontmatter@RRAPformat}
  {}{}
}%
\makeatother
\begin{document}

\preprint{AIP/123-QED}

\title{Quantum smoothed particle hydrodynamics algorithm inspired by quantum walks}

\author{R. Au-Yeung}
\affiliation{Physics Department, University of Strathclyde, Glasgow G4 0NG, UK}
\author{V. M. Kendon}
\affiliation{Physics Department, University of Strathclyde, Glasgow G4 0NG, UK}
\author{S. J. Lind}
\affiliation{School of Engineering, Cardiff University, Cardiff CF24 3AA, UK}

\date{\today}

\begin{abstract}
Recent years have seen great progress in quantum computing, providing opportunities to overcome computational bottlenecks in many scientific applications. In particular, the intersection of computational fluid dynamics (CFD) and quantum computing has become an active area of research with exponential computational speedup as an ultimate goal. In this work, we propose a quantum algorithm for the time-dependent smoothed particle hydrodynamics (SPH) method. Our algorithm uses concepts from discrete-time quantum walks to solve the one-dimensional advection partial differential equation via an SPH formalism. Hence, we construct a quantum circuit to carry out the calculations for a two-particle system over one, two and three timesteps. We compare its outputs with results from the classical SPH algorithm and show there is excellent agreement. The methodology and findings here are a key step towards developing a more general quantum SPH algorithm for solving practical engineering problems on gate-based quantum computers. 
\end{abstract}

\maketitle

\section{Introduction}

Computational fluid dynamics (CFD) simulations are widely used in the automotive, aerospace, civil engineering, renewable energy, and defense industries. These applications typically rely on large-scale numerical simulations to solve the Navier-Stokes equations, running on millions of CPU cores at petaflop speeds. 
We are reaching the limits of what we can do with silicon chip technology and the available power for the largest high performance computing facilities.
It has become clear that we need to develop new methods to perform larger and more complicated computations. Quantum computing is a particularly promising candidate for a range of computational problems. There is evidence it can surpass the most powerful high-performance computers (HPC) \cite{Arute2019,Wu2021,Madsen2022}, when more advanced quantum hardware has been engineered. 
CFD is well-placed to benefit from advances in quantum computing \cite{Li2025,Gaitan2021,Dalzell2025} and the first steps are being taken to develop suitable quantum algorithms. 

Quantum CFD algorithms fit into two broad categories. First, hybrid quantum-classical algorithms \cite{Bharadwaj2023,Jaksch2023,Succi2024} directly solve the equations of motion by outsourcing the parallelizable operations (e.g., solving linear systems \cite{Harrow2009}) to the quantum computer \cite{Oz2021,Jaksch2023,Lapworth2022}. 
Hybrid algorithms may be suitable for the current generation of noisy intermediate-scale quantum (NISQ) computers \cite{Preskill2018,Bharti2022}. 
Bottlenecks in hybrid methods occur during the frequent data exchanges between classical and quantum computers -- the encoding and read-out processes can be more time-consuming than the algorithm itself \cite{Succi2023,Aaronson2015}. 
In the second category, Hamiltonian simulation \cite{Kim2023,Georgescu2014} is better suited for fault-tolerant quantum computers. This method maps the fluid to a quantum system which evolves on the quantum processor. It does not require intermediate state measurements or re-initialization steps. For example, Hamiltonian simulation underpins the quantum lattice Boltzmann method (QLBM) developed by \citeauthor{Succi2015}\cite{Succi2015,Itani2024,Sanavio2024} and \citeauthor{Budinski2021}\cite{Budinski2021}.

Quantum versions of random walks \cite{Aharonov1992,Farhi1998} can be used for building powerful quantum algorithms \cite{Childs2003}.  They have already been used to develop quantum lattice Boltzmann schemes, which have been shown to be formally equivalent to quantum walks \cite{Succi2015}.
Here, we use a quantum walk-based algorithm for the smoothed particle hydrodynamics (SPH) method \cite{Lucy1977,Gingold1977}. 

In a previous work, we presented a proof-of-concept quantum SPH algorithm \cite{AuYeung2024} for solving the one-dimensional advection and diffusion partial differential equations. Now we address a major computational bottleneck in that algorithm: rather than performing quantum encoding and readout at each timestep, we explore how techniques from discrete-time quantum walks can generate multiple timesteps on a quantum computer. This would make the quantum SPH algorithm more efficient by encoding the SPH parameters into the quantum computer, calculating several timesteps, then reading out the current state.  We can repeat this process for longer simulations, with each readout providing a snapshot for data analysis.

This paper is intended for researchers in the fluid mechanics community who may have a limited background in quantum computing. We recommend  \citeauthor{Bharadwaj2023r}'s lecture notes \cite{Bharadwaj2020,Bharadwaj2023r} for a concise overview written from the perspective of CFD applications. Reviews by \citeauthor{Givi2020} \cite{Givi2020} and \citeauthor{Succi2023} \cite{Succi2023} also discuss the challenges facing the quantum CFD field. \citeauthor{Nielsen2010} \cite{Nielsen2010} offer a more pedagogical introduction to quantum computing for further reference.

We want to point out some important differences between quantum and classical computing that appears in our work. A key postulate in quantum mechanics is that all quantum operations must be unitary, linear and reversible. Quantum gates in quantum computers are implemented by unitary operators acting on one or more qubits. Unitary operators $\hat{U}$ have the property that $\hat{U}^{-1} = \hat{U}^\dagger$ where $\hat{U}^{-1}$ performs the inverse operation. Since $\hat{U}^\dagger$ is well defined, the inverse operation must exist. Hence, quantum gates must be reversible. This presents a challenge for nonlinear CFD problems, such as when calculating the nonlinear transformation $u \to u^2$ (velocity $u$) for nonlinear flows. The treatment of non-linearity in quantum computing remains an open question.  Hence, we choose the linear advection equation as the application case in this paper.

The structure of our paper is as follows. Sec. \ref{sec:background} describes the different components required in building our quantum algorithm. We describe the core ideas of SPH in Sec. \ref{sec:coreSPH}, and summarize the novelty in our previous work \cite{AuYeung2024} in Sec. \ref{sec:qsph}. This involves converting the SPH formalism into expressions more suitable for quantum computers. Next, we describe how we adapt concepts from the quantum walk formalism (Sec. \ref{sec:dtqw}) into a quantum smoothed particle hydrodynamics (QSPH) algorithm (Sec. \ref{sec:method}). We provide a fully worked out example using a simple two-particle SPH system (Sec. \ref{sec:2particle}), including numerical simulations using Qiskit software to build the quantum circuit. Then we discuss the results (Sec. \ref{sec:results}) and future work (Sec. \ref{sec:conclusions-future-work}).

\section{Background}\label{sec:background}

This section gives an overview of the SPH algorithm (Sec. \ref{sec:coreSPH}), our previous work on building a quantum SPH algorithm (Sec. \ref{sec:qsph}), and we describe important concepts in the quantum walk formalism (Sec. \ref{sec:dtqw}). We focus on the discrete-time quantum walk on a line, its coin and shift operations, and how we use them in the quantum SPH algorithm in this paper.

\subsection{SPH core concepts}\label{sec:coreSPH}

Ever since SPH was first developed in \citeyear{Lucy1977} for astrophysics simulations \cite{Lucy1977,Gingold1977}, it has been refined and adapted to solve numerous other problems in science and engineering \cite{Monaghan2012}, and more recently in fluid animations for computer graphics applications \cite{Koschier2022}. SPH is a Lagrangian method based on particle interpolation to calculate smooth field variables. These particles carry the physical properties of the system which we update at each timestep. Because of its Lagrangian particle nature, SPH has certain advantages over traditional mesh-based methods. For example, SPH is generally more robust in highly deforming flows and does not suffer from mesh distortions that catastrophically affect the numerical accuracy and stability in mesh-based simulations. See \citeauthor{Monaghan2005}\cite{Monaghan2005} and \citeauthor{Lind2020}\cite{Lind2020} for an introduction and review of the SPH method.

\begin{figure}[ht!]
\centering
\includegraphics[width=\linewidth]{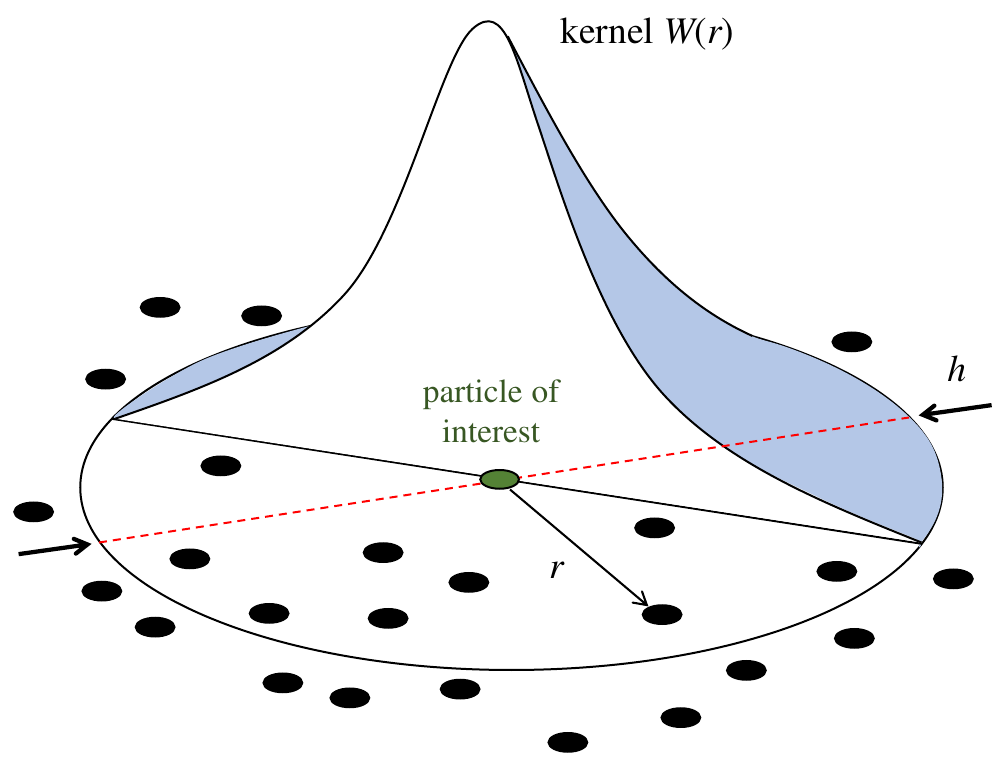}
\caption{Example of a kernel function $W(r = \vert x - y \vert)$ with smoothing length $h$. Support length is equal to the smoothing length in our work, although it is generally integer multiples of $h$. Adapted from \citeauthor{Abaqus}\cite{Abaqus}.}
\label{fig:kernel}
\end{figure}

The foundation of SPH is interpolation theory, based on the Dirac sifting property $f(x) = \int_\Gamma f(y) \delta(x-y) dy$ for some smooth function $f(x)$ and volume of the integral $\Gamma$ that contains $x$. We take the kernel approximation by substituting the Dirac delta distribution with an interpolation function such that
\begin{equation}
f(x) = \int_\Gamma f(y) W(x-y,h) dy.
\end{equation}
The smoothing length $h$ defines the influence (support area) of the smoothing kernel $W$. The kernel must satisfy certain conditions, such as normalization and recovery of the delta function in the limit as $h$ decreases to zero. Other conditions can also be imposed, e.g. compact support \cite{Monaghan2005}. See Fig. \ref{fig:kernel} for an example sketch.

SPH relies on the kernel functions to find the derivatives of continuous fields in a discrete form. This is achieved by converting the continuous interpolation into a particle-based discretization. The resulting equations comprise a discrete mechanical system where particle interactions depend on their mutual distances and mechanical (and possibly thermodynamic) properties. As a result, SPH is consistent with both Lagrangian and Hamiltonian mechanics. This provides a further connection to quantum formalisms and quantum computing \cite{AuYeung2024}. Note that in this work, we consider the Eulerian framework and fix the SPH particle positions so that they do not move in a Lagrangian manner. Not only is Eulerian SPH a valid computational tool in its own right \cite{lind2016high,nasar2019eulerian,nasar2021high}, the approach allows us to first focus on developing a fully quantum time-marching algorithm. We can also generalize to other methods, e.g. finite differences. Subsequent studies will then detail the process of moving and tracking particles in a quantum computing framework which will likely require additional quantum registers. We outline this further work in Sec. \ref{sec:conclusions-future-work}.

\subsection{Summary of previous QSPH work}\label{sec:qsph}

Our previous work \cite{AuYeung2024} presents a proof-of-concept for solving the one-dimensional linear advection equation. It used quantum registers for the spatial derivative but with classical (Eulerian) explicit first-order time-stepping. We will follow the same prescription in the present work, but with the time-stepping now done in a fully quantum setting.  This advection equation can be written as
\begin{equation}
\frac{\partial u(x,t)}{\partial t} + c \frac{\partial u(x,t)}{\partial x} = 0
\end{equation}
with advection quantity $u(x,t)$ as a function of position $x$ and time $t$, and advection speed $c$. This equation can describe a one-dimensional soliton (for example) that moves along a line without change of form. In the SPH formalism, we can rewrite this as
\begin{equation}\label{eq:advection}
u_j(t+\Delta t) = u_j(t) - c \Delta t \sum_k (u_k(t)-u_j(t)) \Delta x_k \nabla_j W_{jk}
\end{equation}
which defines the solutions $u$ of SPH particle $j$ in terms of neighbors $k$ at time intervals defined by timestep size $\Delta t$, with particle spacing $\Delta x$ and first derivative of smoothing kernel $W(r_j-r_k,h)$. Note that we use the zeroth-order consistent formulation for the first derivative.

The crux of our work \cite{AuYeung2024} was to rewrite Eq. \eqref{eq:advection} into a quantum mechanical formalism, 
\begin{equation}\label{eq:q-advection}
u_j(t+\Delta t) = u_j(t) - c \Delta t \nu N \|\vec{a}\| \Re\langle a \vert \nabla_j W_{jk} \rangle
\end{equation}
with normalization constant $\nu = \max \vert \nabla_j W_{jk} \vert$ and $N$ neighbor particles inside region of compact support. We define an inner product $\langle a \vert \nabla_j W_{jk} \rangle$ that contains quantum states
\begin{gather}
\langle a \vert = \frac{\vec{a}^*}{\|\vec{a}\|}, 
\quad 
\vec{a}^* = 
\begin{bmatrix}
(u_1(t)-u_j(t)) \Delta x_1 \\
(u_2(t)-u_j(t)) \Delta x_2 \\
\vdots \\
(u_N(t)-u_j(t)) \Delta x_N
\end{bmatrix} \label{eq:a-state}
\\
\vert \nabla_j W_{jk} \rangle = 
\begin{bmatrix}
\nabla_j W_{j1}/(\nu N) + ib_{j1} \\
\nabla_j W_{j2}/(\nu N) + ib_{j2} \\
\vdots \\
\nabla_i W_{jN}/(\nu N) + ib_{jN}
\end{bmatrix} \label{eq:kernel-state}
\end{gather}
where 
\begin{equation}\label{eq:anorm}
\|\vec{a}\| = \frac{1}{\sqrt{N}} \left( \int_A^B \vert u_k(t)-u_j(t) \vert^2 dx \right)^{1/2}.
\end{equation}

We also introduce constant $b$ to satisfy the normalization conditions of quantum states,
\begin{equation}
b_{jk} = \sqrt{ \frac{1}{N} - \left(\frac{\nabla_j W_{jk}}{\nu N}\right)^2 }
\end{equation}
to ensure the largest absolute value is
\begin{equation}
\bigg\vert \frac{\nabla_j W_{jk}}{\nu N} + ib_{jk} \bigg\vert^2 = 
\frac{1}{N}.
\end{equation}

Our notation essentially recasts the summation [Eq. \eqref{eq:advection}] into an inner product [Eq. \eqref{eq:q-advection}]. The latter can be calculated efficiently on a quantum processor using the swap test \cite{Barenco1997,Buhrman2001}, for example.

It is possible to use quantum encoding procedures to load Eq. \eqref{eq:q-advection} into the quantum computer. For example, quantum amplitude encoding \cite{Long2001,Mottonen2005} can load Eqs. \eqref{eq:a-state} and \eqref{eq:kernel-state} by storing the information in the amplitudes of the quantum state. Then we may use various quantum algorithms to calculate the inner product, such as the swap test \cite{Barenco1997,Buhrman2001} or one of its variants \cite{Fanizza2020}. In the previous work, we pass the solution back to the classical computer and repeat the process to calculate the next timestep.  This is likely to be costly and inefficient.  In this work we fix this problem by extending the quantum calculation to include several timesteps.

We emphasize that this work was a proof-of-concept. We did not determine whether there is any appreciable quantum speed up. Our work solved a one-dimensional advection model to show how a quantum SPH algorithm could work (in theory), and pointed out the numerous issues that must be addressed. We also took many simplifications, such as fixing the SPH particles on a line with equal spacing. This is similar to mesh-based methods: this one-dimensional SPH system is analogous to a finite difference scheme. The inherent strength of SPH is its freely moving particles, so this extra degree of freedom should be considered in future work. Extending the method to two- or three-dimensions is another priority which would make our method more suitable for solving real-world applications. However, we focus on developing a time-stepping mechanism in a quantum framework, inspired by discrete-time quantum walks.

\subsection{Overview of quantum walks}\label{sec:dtqw}

Quantum walks (QWs)\cite{Aharonov1992,Farhi1998} are important theoretical models for quantum computing \cite{Ambainis2003,VenegasAndraca2012}. They form the basis of many quantum algorithms and applications \cite{Kadian2021}, such as simulating complicated fluid flows \cite{Claudon2025,Steijl2018,Zylberman2022}. QWs use quantum superposition and entanglement to provide a more powerful form of classical random walks, as the quantum walker can exist in a superposition of states. QWs are a universal quantum computation primitive \cite{Childs2009} and have been shown to give an exponential algorithmic speed-up \cite{Childs2003}.

QWs come in two flavors: continuous and discrete \cite{Childs2010}. Discrete-time quantum walks (DTQWs) evolve through a sequence of coin operations that determines the quantum walker's direction of movement in position space. In continuous-time quantum walks (CTQWs), the state evolves continuously in time under a Hamiltonian defined by the graph of the position space. 
In this work, we use the DTQW on a line. It is realized by two procedures: the coin operator (to determine the direction of the walk) and shift operator (to transition to the new state determined by the coin). This model uses a qubit coin in $\mathcal{H}_C$ Hilbert space and a set of position states in $\mathcal{H}_P$. The total Hilbert space is denoted $\mathcal{H} = \mathcal{H}_P \otimes \mathcal{H}_C$. The quantum walker evolves according to the unitary operator 
\begin{equation}\label{eq:quantum-walker-operator}
\hat{U}=\hat{S}(\mathbb{1}\otimes\hat{C})
\end{equation}
with coin operator $\hat{C}$ and conditional shift operator $\hat{S}$. 

For the one-dimensional line example, the two-dimensional Hilbert space associated with the coin operator lets the walker choose two possible directions (left or right). Hence, in the unbiased case, the coin is a Hadamard operator,
\begin{equation}
\hat{H}_c = \frac{1}{\sqrt{2}} \begin{bmatrix} 1 & 1 \\ 1 & -1 \end{bmatrix}.
\end{equation}
The shift operator that produces the transition $\ket{j} \to \ket{j \pm 1}$ is
\begin{equation}
\hat{S}_c = \ket{0}\bra{0} \otimes \sum_j\ket{j+1}\bra{j} + \ket{1}\bra{1} \otimes \sum_j\ket{j-1}\bra{j}.
\end{equation}
Hence, each step made by the quantum walker corresponds to the unitary operation $\hat{U}_c=\hat{S}_c(\mathbb{1}\otimes\hat{H}_c)$. As we will explain in Sec. \ref{sec:method}, one QW step is analogous to one timestep in the quantum SPH algorithm.

\section{Method}\label{sec:method}

\begin{figure}[ht!]
\centering
\includegraphics[width=\linewidth]{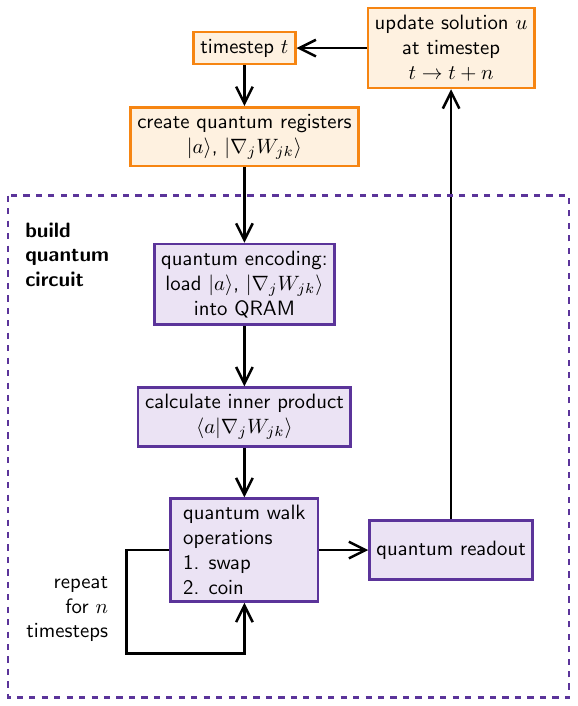}
\caption{Schematics of quantum algorithm. Classical (quantum) procedures in orange (purple). Quantum register $\ket{a}$ contains advection velocities [Eq. \eqref{eq:a-state}]. Other register contains kernel information $\ket{\nabla_j W_{jk}}$ [Eq. \eqref{eq:kernel-state}]. Quantum algorithm encodes SPH parameters into quantum register, then calculates inner product using swap and unitary coin operations.  These gates can be repeated for each timestep iteration before we pass results back to classical computer. \label{fig:flowchart}}
\end{figure}

We provide a high-level summary of the QSPH algorithm in Fig. \ref{fig:flowchart}. The idea is to solve the advection equation over several time steps. We construct a quantum circuit using gates that are analogous to quantum walk operations, namely the swap and coin operators. 
For convenience and to mirror notation in quantum walks, let $\Delta t=1$ in Eq. \eqref{eq:q-advection}. This gives an advection equation
\begin{align}
u_j(t+1) = u_j(t) &\Big(1+ c \nu N \Re \sum_k \Delta x_k V_{jk} \Big) \nonumber\\ 
&- c \nu N \Re \sum_k \Delta x_k u_k(t) V_{jk}
\end{align}
where $V_{jk} = \nabla_j W_{jk}/(\nu N)+ib_{jk}$, $\sum_k$ sums over neighbors $k$ only and implies $k \neq j$. Of course, numerical studies typically require $\Delta t \ll 1$, but, equivalently, we may use $\Delta t=1$ and instead adjust the advection speed to fulfill the Courant-Friedrichs-Lewy (CFL) condition,
\begin{equation}\label{eq:cfl}
\Delta t \leq \frac{\Delta x}{c}.
\end{equation}

For simplicity, we let $\Delta x_k=\Delta x$ so that
\begin{align}
u_j(t+1) = u_j(t) &\Big(1+ c \nu N \Delta x \Re \sum_k V_{jk} \Big) \nonumber\\
&- c \nu N \Delta x \Re \sum_k u_k(t) V_{jk}.
\end{align}

In the QW formalism, we define the amplitudes
\begin{gather}
\alpha_{jj} = 1 + c \nu N \Delta x \Re \sum_k V_{jk}
\label{eq:amplitudes1}\\ 
\alpha_{kj} = - c \nu N \Delta x \Re (V_{jk}),
\label{eq:amplitudes2}
\end{gather}
hence
\begin{equation}\label{eq:u-quantum}
u_j(t+1) = \alpha_{jj} u_j(t) + \sum_k \alpha_{kj} u_k(t).
\end{equation}

Let there be $M$ sites (nodes) in our QW system, analogous to number of SPH particles. There are $N$ neighbors per site, and $N<M$ with $N$ neighbors for each site.

Next, we store scalars $u_j(t)$ in vector $\vec{u}(t)$ and use quantum amplitude encoding to create a register
\begin{equation}
\ket{\vec{u}(t)} = \sum_{j=1}^{M} u_j(t) \ket{j}
\end{equation}
in computational basis $\ket{j}$ containing particle positions $j$. 
Summation over all neighbors is the most computationally expensive process that we can possible do. Hence for each site $j$, we want to select a subset of size $N$ neighbors (analogous to the kernel function selecting neighbors within its support), multiply $u$ with corresponding amplitudes $\alpha$, and finally perform the summation.

We need another register to label the neighbors. At each site, we define vector $\vec{u}_j(t)$ containing $N+1$ components for $N$ neighbors with site indices $k$, $k'$, $k''$, ... to produce a vector
\begin{equation}\label{eq:eq11}
\vec{u}_j(t) = 
\begin{bmatrix} u_{jj}(t) \\ u_{jk}(t) \\ u_{jk'}(t) \\ \vdots
\end{bmatrix}.
\end{equation}
Then we encode the neighbor information by including a second register,
\begin{equation}
\ket{\vec{u}(t)} = \sum_{j=1}^{M}\sum_{n}^{N} u_{jn}(t) \ket{j,n}
\end{equation}
where $n$ is an element of the neighbors subset. Note we use notation $\ket{j,k}$ to represent two separate registers, where $j$ encodes the particle positions and $k$ for the neighbors. This is equivalent to QW position and coin states.

We apply the unitary shift operator $\hat{S}$,
\begin{equation}
\hat{S}\ket{\vec{u}(t)} = \sum_{j=1}^{M}\sum_{n}^{N} u_{nj}(t) \ket{j,n}
\end{equation}
to reorganize the neighbors so that for site $j$,
\begin{equation}
\vec{u}_j(t) = 
\begin{bmatrix} u_{jj}(t) \\ u_{kj}(t) \\ u_{k'j}(t) \\ \vdots
\end{bmatrix}.
\end{equation}
Reordering the vector corresponds to a QW propagation operation. It transfers the neighbor information to particle $j$. This is needed for calculating the kernel interaction. Therefore we need to solve 
\begin{equation}\label{eq:solve_u}
u_j(t+1) = \alpha_{jj} u_j(t) + \sum_n \alpha_{nj} u_n(t) = \vec{u}_j(t) \cdot \vec{\alpha}_j
\end{equation}
where vector $\vec{\alpha}_j$ contains $\alpha_{jk}$ terms. Because this needs to be the same size as $\ket{\vec{u}(t)}$ to calculate an inner product, we let
\begin{equation}
\ket{\vec{\alpha}} = \sum_j \ket{\vec{\alpha}_j} = \sum_{j,n}\alpha_{nj} \ket{j,n}.
\end{equation}

The inner product is $\vec{u}_j(t) \cdot \vec{\alpha}_j \to \langle\vec{u}(t)\vert\vec{\alpha}\rangle$, where it is generally inefficient to prepare the $\ket{\vec{\alpha}}$ state from a classical description. Quantum algorithms typically calculate an inner product as final measurement. It is an irreversible procedure and therefore a non-unitary operation. 
In our previous work\cite{AuYeung2024}, we outlined several methods for efficiently calculating the inner product to output a classical value.
However, to compute several timesteps within the quantum part of the algorithm, we need use a reversible unitary operation. We model this on the ``coin operation'' from quantum walks.  This accomplishes the summation, but also produces extra ``junk'' terms $*$ that contain the information required to reverse the operation,
\begin{equation}
\vec{u}_j(t+1) = 
\begin{bmatrix}
u_j(t+1) \\ * \\ * \\ * \\ \vdots
\end{bmatrix}.
\end{equation}
However, the $*$ terms reduce the probability of measuring the correct output. There are several options to reduce or remove the $*$ entries. 
For example, we can use quantum amplitude amplification \cite{Brassard1997,Grover1998} to maximize $u_j(t+1)$ and minimize the $*$ terms, then discard the qubits carrying the $*$ information. Another option involves the ``uncomputation trick'' to disentangle the neighbor register \cite{Bennett1997,Cleve2013}.
In all cases, we need to use extra qubits and gate operations to proceed to the next timestep.  In our work, we leave the $*$ terms associated with the old neighbor register, add a new neighbor register, entangle it with the site register, then calculate the next timestep.  We then post-select for the result we want at the end of the computation.  This lowers the overall success probability, but simplifies the presentation of the method in the simple example in Sec. \ref{sec:2particle}.

\section{Simulation of two-particle system}\label{sec:2particle}

\begin{figure}[ht!]
\centering
\includegraphics[width=\linewidth]{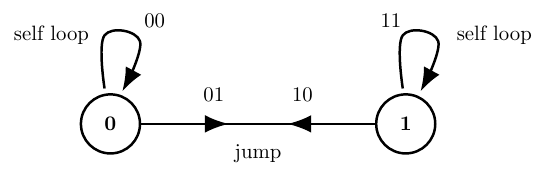}
\caption{Example system with two sites (SPH particles) only.}
\label{fig:2particles}
\end{figure}

In this section, we present a fully worked out example using a simple two-particle model (Fig. \ref{fig:2particles}). We consider two sites labeled ``0'' and ``1''. They respectively correspond to states $u_0$ and $u_1$. The quantum walker can initially begin at site 0 (1), then either jump to site 1 (0) or stay at site 0 (1).

\begin{figure}[ht!]
\centering
\includegraphics[width=\linewidth]{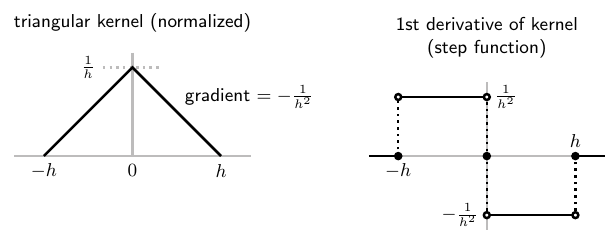}
\caption{Sketch of triangular kernel function (with smoothing length $h$) and its first derivative.}
\label{fig:tri-kernel}
\end{figure}

For a two-particle system, it is suitable to use a triangular smoothing kernel (Fig. \ref{fig:tri-kernel}) of the form
\begin{align}
W(r, h) &=
\begin{cases}
1/h - \vert r \vert / h^2, & \vert r \vert < h \\
0, & \text{otherwise} 
\end{cases}, \\
\frac{dW(r, h)}{dr} &=
\begin{cases}
-\text{sgn}(r)/h^2, & \vert r \vert < h \\
0, & \text{otherwise} 
\end{cases}.
\end{align}
Given only two particles, this simple kernel structure essentially reproduces a first-order finite difference approximation for the spatial derivative. There is little merit in using higher-order (i.e. smoother, Gaussian-type kernels) kernels over this two-particle system. Any benefits smoother kernels may have on accuracy and stability are only apparent for many-particle systems.

\subsection{Quantum registers}

Let the system wavefunction at time $t$ be
\begin{align}
\psi(t) &= \beta_{00}(t)\ket{0,0} + \beta_{01}(t)\ket{0,1} \nonumber\\
&+ \beta_{10}(t)\ket{1,0} + \beta_{11}(t)\ket{1,1}
\end{align}
for arbitrary amplitudes $\beta_{k\ell}$. The subscripts $k$ and $\ell$ represent the quantum walker's initial and final position respectively. The quantum state $\ket{\text{position},\text{coin}}$ contains information on the quantum walker position and coin operator.

The ``velocity'' register contains information on the SPH particle velocity, which we define as
\begin{equation}\label{eq:uket}
\ket{\vec{u}(t)} = u_0(t)\ket{0} + u_1(t)\ket{1} = 
\mathcal{C} \begin{bmatrix} u_0(t) \\ u_1(t) \end{bmatrix}
\end{equation}
with normalization constant $\mathcal{C}$. The neighbor register 
\begin{equation}\label{eq:neighbor-register}
\ket{v} = \frac{1}{\sqrt{2}} (\ket{0}+\ket{1})
\end{equation}
is a normalized state containing all SPH particles inside the region of compact support. 
(Note that for brevity, we will use the term ``velocity register'' to refer to the advection quantities $u$ which are solutions to the advection differential equation.)
The tensor product gives entangled state
\begin{align}
\psi(t) 
&= \ket{\vec{u}(t)} \otimes \ket{v} \\
&= \frac{\mathcal{C}}{\sqrt{2}} ( u_0(t)(\ket{0,0}+\ket{0,1}) + u_1(t)(\ket{1,0}+\ket{1,1}) ) \label{eq:ket} \\
&= \frac{\mathcal{C}}{\sqrt{2}} \begin{bmatrix} u_0 \\ u_0 \\ u_1 \\ u_1 \end{bmatrix}
\end{align}
such that $\beta_{00}=\beta_{01}=\mathcal{C}u_0/\sqrt{2}$ and $\beta_{10}=\beta_{11}=\mathcal{C}u_1/\sqrt{2}$.

\subsection{Shift operation}

Applying the shift operator 
\begin{gather}\label{eq:shift}
\hat{S} = \begin{bmatrix} 1 & 0 & 0 & 0 \\ 0 & 0 & 1 & 0 \\ 0 & 1 & 0 & 0 \\ 0 & 0 & 0 & 1 \end{bmatrix} 
\end{gather}
to the entangled state gives
\begin{gather}
\hat{S}\psi(t) = \frac{\mathcal{C}}{\sqrt{2}} \begin{bmatrix} u_0 \\ u_1 \\ u_0 \\ u_1 \end{bmatrix}.
\end{gather}
This is analogous to the DTQW shift operation and transfers neighbor information between the SPH particles.

\subsection{Coin}

Now use Eq. \eqref{eq:solve_u} to define the next timestep,
\begin{align}
\begin{bmatrix} u_0(t+1) \\ u_1(t+1) \end{bmatrix}
&=
\begin{bmatrix} 
\alpha_{00} u_0(t) + \alpha_{10} u_1(t) \\ 
\alpha_{11} u_1(t) + \alpha_{01} u_0(t) 
\end{bmatrix}
\\
&=
\begin{bmatrix} 
\alpha_{00} & \alpha_{10} \\ 
\alpha_{01} & \alpha_{11}
\end{bmatrix}
\begin{bmatrix} 
u_0(t) \\ u_1(t)
\end{bmatrix}.
\end{align}

As described above, we need to include extra ``junk" terms and expand the state space so that we can do a ``reversible'' inner product operation. The matrix would have the structure:
\begin{equation}
\begin{bmatrix} u_0(t+1) \\ * \\ * \\ u_1(t+1) \end{bmatrix}
=
\begin{bmatrix} 
\begin{matrix}
\alpha_{00} & \alpha_{10} \\ * & *
\end{matrix}
& \rvline & \bigzero \\
\hline
\bigzero & \rvline &
\begin{matrix}
* & * \\ \alpha_{01} & \alpha_{11}
\end{matrix}
\end{bmatrix}
\begin{bmatrix} 
u_0(t) \\ u_1(t) \\ u_0(t) \\ u_1(t)
\end{bmatrix}
\end{equation}
which contains amplitudes $\alpha$ defined in Eqs. \eqref{eq:amplitudes1} and \eqref{eq:amplitudes2}, and unwanted terms $*$.  Since the $2\times2$ sub-blocks must be unitary, we can let
\begin{align}\label{eq:unext}
\begin{bmatrix} u_0(t+1) \\ * \\ * \\ u_1(t+1) \end{bmatrix}
&=
\begin{bmatrix} 
\begin{matrix}
\alpha_{00} & \alpha_{10} \\ \alpha_{10} & - \alpha_{00}
\end{matrix}
& \rvline & \bigzero \\
\hline
\bigzero & \rvline &
\begin{matrix}
-\alpha_{11} & \alpha_{01} \\ \alpha_{01} & \alpha_{11}
\end{matrix}
\end{bmatrix}
\begin{bmatrix} 
u_0(t) \\ u_1(t) \\ u_0(t) \\ u_1(t) 
\end{bmatrix}
\\
&=
\begin{bmatrix} 
\alpha_{00} u_0(t) + \alpha_{10} u_1(t) \\ 
\alpha_{10} u_0(t) - \alpha_{00} u_1(t) \\ 
\alpha_{01} u_1(t) - \alpha_{11} u_0(t) \\ 
\alpha_{11} u_1(t) + \alpha_{01} u_0(t) 
\end{bmatrix}\label{eq:final-state}
\end{align}

This step is analogous to applying the coin operation,
\begin{equation}
\mathbb{1}\otimes\hat{C} = 
\begin{bmatrix} \hat{H}_0 & 0 \\ 0 & \hat{H}_1 \end{bmatrix}
\end{equation}
where we define the coins
\begin{gather}
\hat{H}_0 = 
\begin{bmatrix}
\alpha_{00} & \alpha_{10} \\ \alpha_{10} & - \alpha_{00}
\end{bmatrix}
\\
\hat{H}_1 = 
\begin{bmatrix}
-\alpha_{11} & \alpha_{01} \\ \alpha_{01} & \alpha_{11}
\end{bmatrix}.
\end{gather}

We simplify the coins using Eqs. \eqref{eq:amplitudes1} and \eqref{eq:amplitudes2}:
\begin{align}\label{eq:alphas}
\alpha_{11} = \alpha_{00}, 
\quad 
\alpha_{10} = \alpha_{01}, 
\quad  
\alpha_{10} = 1-\alpha_{00}.
\end{align}
Hence
\begin{gather}\label{eq:coinmatrix}
\mathbb{1}\otimes\hat{C} =
\begin{bmatrix}
\alpha_{00} & 1-\alpha_{00} & 0 & 0 \\ 
1-\alpha_{00} & -\alpha_{00} & 0 & 0 \\
0 & 0 & -\alpha_{00} & 1-\alpha_{00} \\ 
0 & 0 & 1-\alpha_{00} & \alpha_{00}
\end{bmatrix}.
\end{gather}

The coin in its current form needs to be normalized. This gives 
\begin{gather}\label{eq:coin}
\hat{U}_C = \mathcal{N} (\mathbb{1}\otimes\hat{C})
\end{gather}
where normalization constant $\mathcal{N} = (\alpha_{00}^2 + (1-\alpha_{00})^2)^{-1/2}$ and $\alpha_{00} = 1 + c \Delta x \nabla_{0} W_{01}$. The $\alpha_{00}$ terms provide a mapping to SPH particles.

\subsection{Quantum circuit}\label{sec:quantum-circuit}

\begin{figure*}[ht!]
\centering
\includegraphics[width=\linewidth]{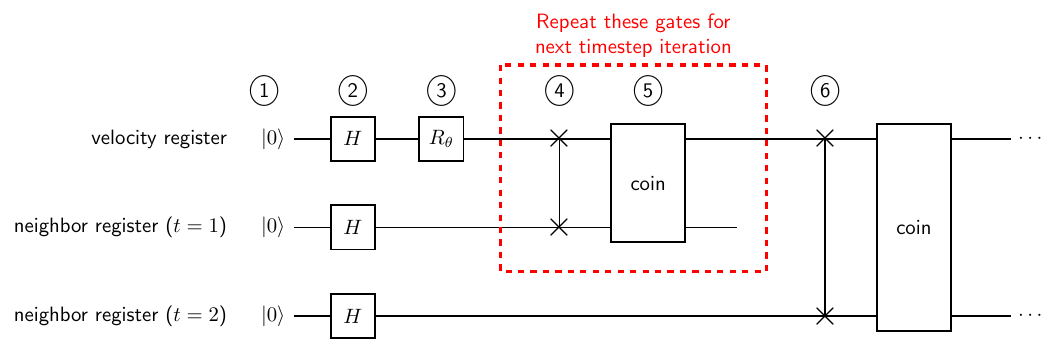}
\caption{Circuit schematics for simulating SPH system containing two particles over multiple timesteps.}
\label{fig:annotated-circuit}
\end{figure*}

Given the initial states $u_0(0)$ and $u_1(0)$, the aim is to simulate the evolution across one timestep to obtain $u_0(1)$ and $u_1(1)$. We construct a quantum circuit to perform the entanglement, shift, and coin operations (Fig. \ref{fig:annotated-circuit}):

\begin{enumerate}[label=\large\protect\textcircled{\small\arabic*}]

\item First, we initialize the system. We need one ``velocity'' register, containing the advection quantity $u$, and one neighbor register. Each contain one qubit. Both are in the ground state $\ket{0}=\begin{bmatrix} 1 \\ 0 \end{bmatrix}$.

\item Hadamard gates create an equal superposition of quantum states, 
\begin{equation}
\ket{\text{velocity}} = \ket{\text{neighbors}} = \frac{1}{\sqrt{2}} \begin{bmatrix} 1 \\ 1 \end{bmatrix}
\end{equation}

\item For the velocity register, we encode the initial condition $[u_0(0), u_1(0)]$ using amplitude encoding \cite{GonzalezConde2024} such that
\begin{equation}
\ket{\text{velocity}} = \mathcal{C} \begin{bmatrix} u_0(0) \\ u_1(0) \end{bmatrix}
\end{equation}
with normalization constant $\mathcal{C} = (u_0(0)^2 + u_1(0)^2)^{-1/2}$. Since there is one qubit to initialize, we may use a rotation matrix to do the encoding. 

As a result, we have an entangled state 
\begin{equation}
\ket{\text{velocity}} \otimes \ket{\text{neighbors}} = 
\frac{\mathcal{C}}{\sqrt{2}}
\begin{bmatrix} u_0(0) \\ u_0(0) \\ u_1(0) \\ u_1(0) \end{bmatrix}.
\end{equation}

\item Next, the swap operation [or shift operator in Eq. \eqref{eq:shift}] produces the state 
\begin{equation}
\frac{\mathcal{C}}{\sqrt{2}}
\begin{bmatrix} u_0(0) \\ u_1(0) \\ u_0(0) \\ u_1(0) \end{bmatrix}.
\end{equation}
Physically, it means the particle information is transferred (swapped) between the two particles. This is analogous to setting up the SPH kernel interaction. 

\item The coin operation [Eq. \eqref{eq:coin}] gives a final state that contains $u_{0,1}(1)$ plus two unwanted terms [Eq. \eqref{eq:final-state}],
\begin{equation}\label{eq:u1-state}
\frac{\mathcal{CN}}{\sqrt{2}}
\begin{bmatrix} 
\alpha_{00} u_0(0) + (1-\alpha_{00}) u_1(0) \\ 
(1-\alpha_{00}) u_0(0) - \alpha_{00} u_1(0) \\ 
(1-\alpha_{00}) u_1(0) - \alpha_{00} u_0(0) \\ 
\alpha_{00} u_1(0) + (1-\alpha_{00}) u_0(0) 
\end{bmatrix}.
\end{equation}

\item To calculate the next timestep, we repeat the previous two gates (swap and coin) using the velocity register and a new neighbor register.

\end{enumerate}

Note that if we want to calculate the time evolution over one timestep, we omit step \textcircled{\small{6}} and all subsequent gate operations. Instead, we would measure the velocity and neighbor qubit registers immediately after step \textcircled{\small{5}}, and extract the solutions $u_{0,1}(1)$. This is shown in Sec. \ref{sec:results}.

\subsection{Results and discussion}\label{sec:results}

\begin{figure}[ht!]
\centering
\includegraphics[width=\linewidth,trim={1cm .5cm 0 1cm},clip]{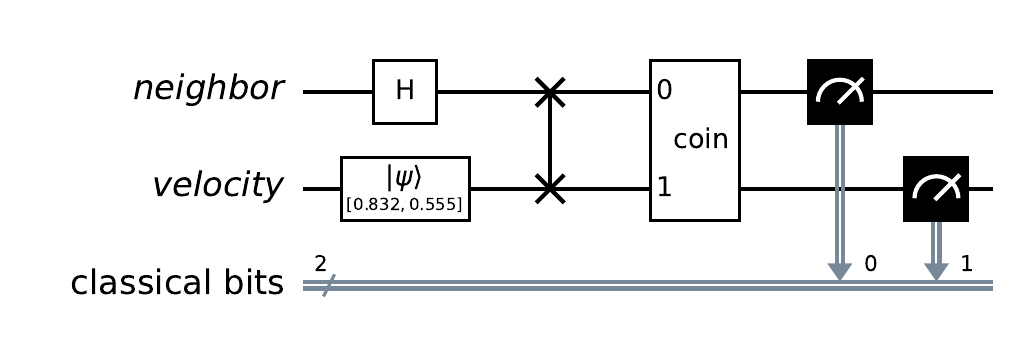}
\caption{Simulating two SPH particles over one timestep. Qiskit circuit performs entanglement, swap, coin operation, disentanglement, then takes final measurement.}
\label{fig:circuit-qiskit}
\end{figure}

We build a circuit using IBM's Qiskit software package (Fig. \ref{fig:circuit-qiskit}) that performs the calculations in Fig. \ref{fig:annotated-circuit} for one timestep. As discussed Sec. \ref{sec:quantum-circuit}, we use two quantum registers labeled ``velocity'' and ``neighbor'' plus two classical bits needed for measuring the final state. Each register contains one qubit in the $\ket{0}$ state. We use Qiskit's state preparation functions to initialize the velocity and neighbor statevectors, where the former contains the initial advection quantities $u_{0,1}(0)$. The CNOT gate creates an entangled state, and the swap gate which transfers neighbor information between the two SPH particles. Next, the unitary coin operator calculates the advection quantities at the next timestep $u_{0,1}(1)$. Finally, we measure the velocity and neighbor qubits to obtain the system statevector. In the post-processing step, we multiply the Qiskit solutions by the normalization constants $(\mathcal{CN}/\sqrt{2})^{-1}$ [see constant in Eq. \eqref{eq:u1-state}].

\begin{figure*}[ht!]
\centering
\includegraphics[width=\linewidth]{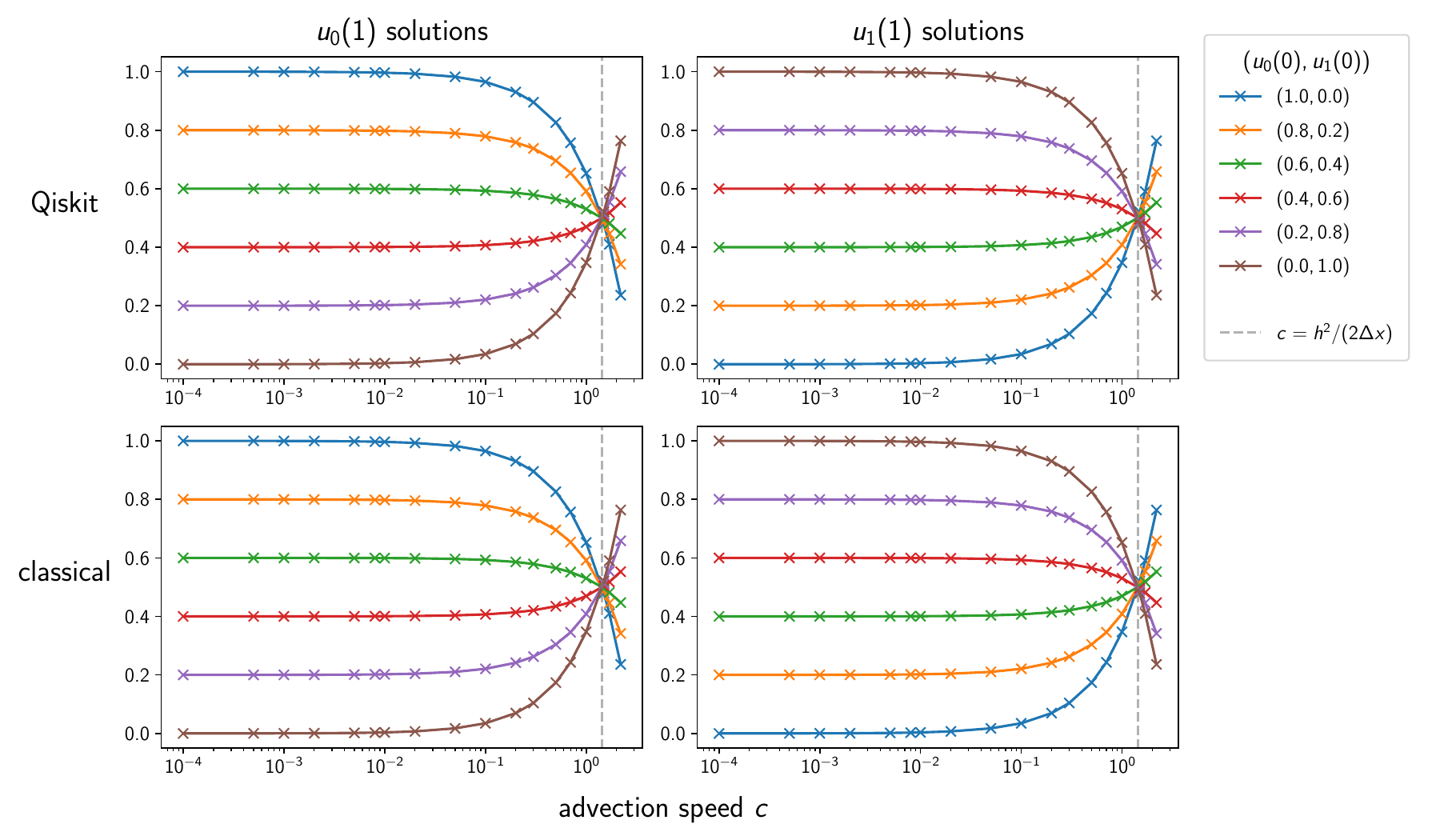}
\caption{Results $u_{0,1}(1)$ after one timestep with initial states $u_{0,1}(0)$, particle separation $\Delta x = 0.5$, kernel smoothing length $h=1.2$, and varying advection speed $c$. Top and bottom rows show Qiskit and classical solutions respectively. Left and right columns show $u_0(1)$ and $u_1(1)$ solutions.}
\label{fig:results1}
\end{figure*}

In Fig. \ref{fig:results1}, we graphically compare the Qiskit and classical solutions [Eq. \eqref{eq:advection}] for which this simplified system reduces to
\begin{equation}
u_{0,1}(1) = \alpha_{00} u_{0,1}(0) + (1-\alpha_{00}) u_{1,0}(0).
\end{equation}
Each row shows the Qiskit and classical solutions, whereas each column shows the solutions of $u_0$ and $u_1$ after one timestep. Each graph color represents a different set of initial conditions $u_{0,1}(0)$. We vary the advection speed $c$ from $10^{-4}$ to $2$. 
For each set of initial conditions in Fig. \ref{fig:results1}, the solutions are $u_0(1)+u_1(1)=1$, indicating a conservation of energy.
For small $c$, the solutions $u_0(1) \approx u_0(0)$ and $u_1(1) \approx u_1(0)$. As $c$ increases, $u_0(1)$ and $u_1(1)$ gradually converge to $(u_0(0)+u_1(0))/2$. For all initial conditions in Fig. \ref{fig:results1}, $(u_0(0)+u_1(0))/2=0.5$. Hence, all graphs converge to 0.5 at some critical advection speed. By setting $u_0(1)=u_1(1)$, we deduce that this critical point occurs at
\begin{align}\label{eq:crossover}
c = -\frac{1}{2} (\Delta t \Delta x \nabla_0 W_{01})^{-1}.
\end{align}

When $\Delta x < h$, $\nabla_0 W_{01} = -1/h^2$ and we can simplify Eq. \eqref{eq:crossover} to
\begin{equation}\label{eq:c-crossover}
c = \frac{h^2}{2 \Delta t \Delta x}, \quad \Delta t = 1.
\end{equation}
This crossover is indicated as gray dashed vertical lines in Fig. \ref{fig:results1}. Given $h \approx \Delta x$, this crossover is approximately half the CFL limit for explicit schemes for the advection equation. Hence, the critical point is consistent with the traveling solution peak ($u_0 + u_1$) occupying a position halfway between the particles. 

\begin{figure*}[ht!]
\centering
\includegraphics[width=\linewidth]{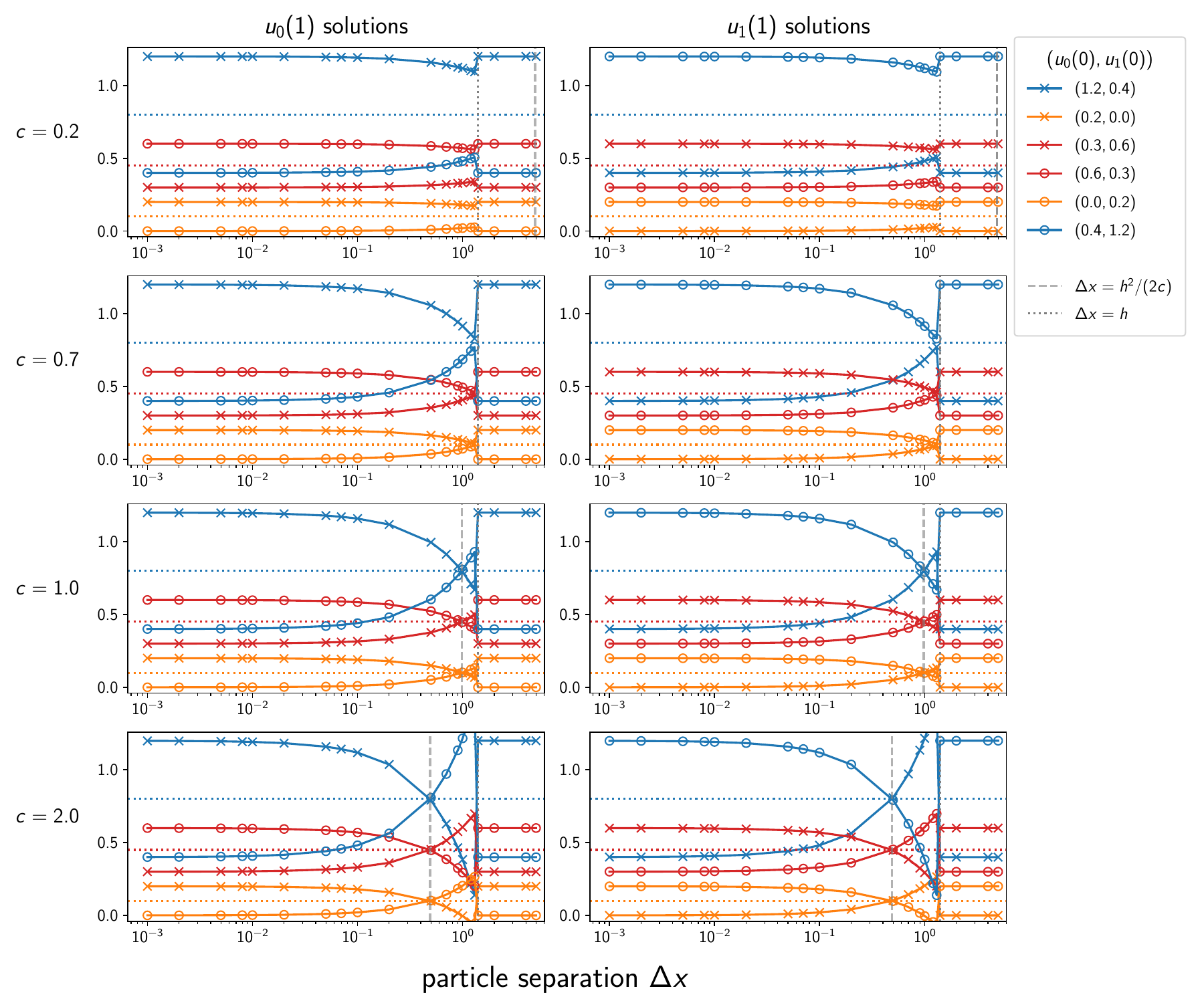}
\caption{Qiskit results $u_{0,1}(1)$ after one timestep with initial states $u_{0,1}(0)$, kernel smoothing length $h=1.4$, and varying particle separation $\Delta x$. Rows show different advection speeds $c$. Left and right columns show $u_0(1)$ and $u_1(1)$ solutions.}
\label{fig:results3}
\end{figure*}

Next, we try a different set of parameters (Fig. \ref{fig:results3}). We vary the particle separation $\Delta x$ and fix the advection speed $c$ and smoothing length $h$. As with the results in Fig. \ref{fig:results1}, the Qiskit and classical solutions agree to machine precision (within $10^{-16}$), so we do not show the classical solutions in Fig. \ref{fig:results3} for clarity. Using Eq. \eqref{eq:crossover}, we expect a crossover to occur when
\begin{equation}\label{eq:Deltax-crossover}
\Delta x = \frac{-1}{2c \Delta t \nabla_0 W_{01}},
\end{equation}
shown as gray dashed vertical lines in Fig. \ref{fig:results3}. Hence, using the kernel property of compact support,
\begin{equation}
\Delta x_\text{crossover} = 
\begin{cases}
h^2 / (2 c \Delta t) & \Delta x < h \\
\text{undefined} & \Delta x \geq h 
\end{cases}.
\end{equation}
When $\Delta x \geq h$, outside the area of compact support, the kernel function has no effect on the SPH particles and $u_{0,1}(0) = u_{0,1}(1)$. It is interesting to note this behavior, even as we explicitly define the neighbor register [Eq. \eqref{eq:neighbor-register}] to contain only the SPH particles inside the area of compact support.

There are some cases where the solutions become negative (Fig. \ref{fig:results3}). When $c=2$ and initial states are $(u_0(0), u_1(0)) = (0.2, 0)$ and $(u_0(0), u_1(0)) = (0, 0.2)$, we see that $u_0(1)<0$ and $u_1(1)<0$ respectively, for larger particle spacing values. Negative solutions indicate numerical instabilities. This behavior is unsurprising at larger $\Delta x$, and since there are only two SPH particles, the system is likely to be prone to instability in any case. We would have a clearer picture on stability behavior once we consider more particles.

\begin{figure*}[ht!]
\centering
\includegraphics[width=\linewidth]{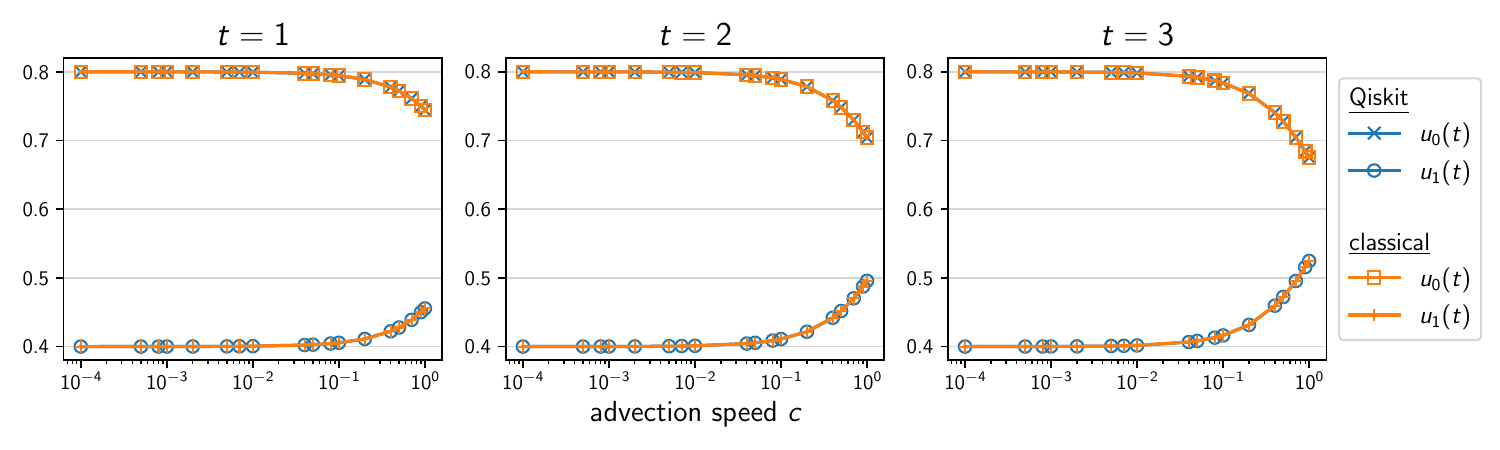}
\caption{Results $u_{0,1}(t)$ after $t=1,2,3$ timesteps with initial states $(u_0(0), u_1(0))=(0.8, 0.4)$, particle separation $\Delta x = 0.2$, kernel smoothing length $h=1.2$, and varying advection speed $c$. Blue and orange graphs show Qiskit and classical solutions respectively.}
\label{fig:results5}
\end{figure*}

Finally, we present the results after several timesteps (Fig. \ref{fig:results5}). Each graph from left to right shows the solutions after $t=1$, $2$, and $3$ timesteps as we vary the advection speed $c$ from $10^{-4}$ to $1$. Qiskit and classical results are in good agreement. 

There are unwanted terms in the statevector after simulating one timestep [Eq. \eqref{eq:u1-state}]. These have a significant effect on the solution accuracy for the $t=2,3$ results. We need to remove the unwanted terms after measurement, in the post-processing stage (see Appendix). Calculating $u_{0,1}(2)$ and $u_{0,1}(3)$ involve taking linear combinations of the statevector elements. This becomes non-trivial when the circuit contains many qubits or if we want to calculate many timesteps. In this case, it will be useful to explore the techniques described at the end of Sec. \ref{sec:method}, such as quantum amplitude amplification or the ``uncomputation trick''. It is essential to address this issue before trying to simulate more timesteps.

\section{Conclusions and future work} \label{sec:conclusions-future-work}

We develop a quantum algorithm based on QW operations as a framework to solve the advection equation in SPH form. We provide a fully worked out example using a simple two-particle system. Then we build a quantum circuit using Qiskit software to perform the simulation over three timesteps. We present several graphs showing the results as we vary the initial advection quantity, advection speed, and SPH particle separation. There is excellent agreement with results from classical calculations. However, due to unwanted terms in the statevector after simulating one timestep, the probability of measuring the solution can deteriorate for further timesteps. This is addressed by removing the unwanted terms after measurement in the post-processing stage.

Our progress sets up numerous future research avenues. For example, we could generalize the procedures in this paper to analyze the advection and diffusion equations containing many SPH particles over many timesteps. Hence, leading to the application end goal of solving the Navier-Stokes equation.

\subsection{Freely moving particles}

We used quantum walk principles to develop a timestep method that is already more general than a basic quantum walk.  The next step is to move the particles off grid.  One of the strengths of SPH over mesh-based methods is its freely moving SPH particles. Including these extra degrees of freedom would align the quantum algorithm closer to classical SPH formalism. This would go beyond quantum walks by introducing a third register for the particle positions which also needs to be updated appropriately.  However, the same basic methods for calculating the updates to each vector can continue to be adapted to the new degrees of freedom.

To calculate the SPH kernel function, we need to determine the particles within the radius of the smoothing length (inside the interaction range). Numerous methods have been developed to efficiently create a list of neighbors, such as the cell-linked list and Verlet list schemes \cite{DominguezAlonso2014}. In the quantum domain, Grover's search algorithm \cite{Grover1996} provides quadratic speedup over classical search methods. Combining Grover search with existing SPH neighbor-list approaches could offer a novel and effective search method.

\subsection{More particles and timesteps}

We can generalize the two-particle example by considering several SPH particles organized into simple geometries. Examples include particles on a line (one-dimensional system) \cite{AuYeung2024} or in a circular arrangement \cite{Wadhia2024}. We note that discrete-time quantum walks on cycle graphs \cite{VenegasAndraca2012,Wadhia2024} have a convenient encoding when executed on digital quantum computers. We can also consider a system of disordered particles with a constrained quantum walk lattice. In any case, this will require more qubits.

To evolve the system over many timesteps, we need more qubits and gates, hence increasing the quantum circuit depth. As mentioned above, one options is to perform quantum amplitude amplification. The aim is to discard the unwanted terms in the statevector. Hence, after performing the coin operation (step \textcircled{\small{5}} in Fig. \ref{fig:annotated-circuit}), amplitude amplification would minimize the $\beta_{01}$ and $\beta_{10}$ terms in Eq. \eqref{eq:u1-state}, while maximizing the $\beta_{00}$ and $\beta_{11}$ terms. 

Quantum algorithms that simulate more SPH particles and timesteps require deep quantum circuits composed of highly accurate quantum gates. We do not expect our algorithm to be viable for noisy intermediate-scale quantum (NISQ) computers \cite{Preskill2018}. Rather, we anticipate our future QSPH algorithms to be run on fully error-corrected, fault-tolerant quantum hardware. In such hardware, error correction \cite{Roffe2019,Terhal2015} and error mitigation techniques \cite{Cai2023} are applied below the level of the algorithm so they are not application-specific, and we can assume we have almost perfect hardware to run our algorithm.

\subsection{Resource requirements}

Although it is important to consider real-world applications containing potentially millions of SPH particles evolving over many timesteps, our focus in this work is on the timestepping. At this stage, it is premature to do large-scale resource estimates before further algorithm testing and refinement.

The SPH method does not require reading out the full information of all SPH particles. Rather, we can use the already smoothed/interpolated function values for a subset of particles to provide a decent representation of the fluid. Since we expect to measure the quantum circuit at regular intervals, the number of timesteps that we run before measurement and reinitialization will be roughly constant, even as the problem size increases.

At the start of the calculation, we encode the initial state into the velocity register (Fig. \ref{fig:annotated-circuit}). For example, quantum amplitude encoding stores the information in the amplitudes of the quantum state. The qubit number scales logarithmically with the input vector length, whereas gate count scales exponentially with the input vector length. This is true when each amplitude is different. If adjacent amplitudes are related because we have a smoothed function, this can be mitigated to generate the SPH particles that represent the function \cite{Hayes2023}.

Finally, we reiterate that the quantum walk formalism has given us the mechanism for calculating a pair-wise particle interaction. This is the fundamental unit of operation on which all SPH methods are built. The underpinning quantum algorithm is valid, regardless of whether the flow is turbulent or multiphase.

\section*{Data Availability Statement}

The data that support the findings of this study are available from the corresponding author upon reasonable request.

\begin{acknowledgments}
RA, SL, and VK are supported by UK Research and Innovation (UKRI) grants EP/Y004566/1 (RA and VK) and EP/Y004663/2 (SL) (QuANDiE) and EP/Z53318X/1 (QCI3). RA and VK are supported by EP/W00772X/2 (QEVEC) and EP/T001062/1 (QCS Hub). VK is supported by EP/T026715/2 (CCP-QC). 

For the purpose of open access, the authors have applied a Creative Commons Attribution (CC BY) licence to any Author Accepted Manuscript version arising from this submission.
\end{acknowledgments}

\appendix

\section*{Author declarations}

\subsection*{Conflict of interest}

The authors have no conflicts to disclose.

\subsection*{Author contributions}

\textbf{RA}: Conceptualization, data curation, formal analysis, funding acquisition, investigation, methodology, project administration, software, supervision, validation, visualization, writing – original draft, writing – review \& editing

\textbf{VK}: Conceptualization, formal analysis, funding acquisition, methodology, project administration, resources, supervision, writing – original draft, writing – review \& editing 

\textbf{SL}: Conceptualization, formal analysis, funding acquisition, methodology, project administration, supervision, writing – original draft, writing – review \& editing

\appendix*

\section{Post-processing Qiskit outputs after two timesteps}

In this section, we calculate the advection solution at $t=2$. We start with the advection solution at $t=1$ [Eq. \eqref{eq:u1-state}, or between step \textcircled{\small{5}} and \textcircled{\small{6}} in Fig. \ref{fig:annotated-circuit}]. At this point, the statevector for the two-qubit system is
\begin{equation}\label{eq:psit1}
\begin{bmatrix} 
\psi_0 \\ \psi_1 \\ \psi_2 \\ \psi_3 
\end{bmatrix}
=
\frac{\mathcal{CN}}{\sqrt{2}}
\begin{bmatrix} 
\alpha_{00} u_0(0) + (1-\alpha_{00}) u_1(0) \\ 
(1-\alpha_{00}) u_0(0) - \alpha_{00} u_1(0) \\ 
(1-\alpha_{00}) u_1(0) - \alpha_{00} u_0(0) \\ 
\alpha_{00} u_1(0) + (1-\alpha_{00}) u_0(0) 
\end{bmatrix}.
\end{equation}

Alternatively, the three-qubit system containing the velocity register qubit and two neighbor register qubits can be expressed as entangled state,
\begin{align}
\begin{bmatrix} 
\psi_0 \\ \psi_1 \\ \psi_2 \\ \psi_3 
\end{bmatrix}
\otimes
\frac{1}{\sqrt{2}}
\begin{bmatrix} 1 \\ 1 \end{bmatrix}
=
\frac{1}{\sqrt{2}}
\begin{bmatrix} 
\psi_0 \\ \psi_0 \\ \psi_1 \\ \psi_1 \\ \psi_2 \\ \psi_2 \\ \psi_3 \\ \psi_3
\end{bmatrix}
\end{align}

Next, applying the swap gate $\hat{S}$ [step \textcircled{\small{5}} in Fig. \ref{fig:annotated-circuit}] gives the statevector
\begin{align}
\frac{1}{\sqrt{2}}
&\underbrace{\begin{bmatrix} 
1 & 0 & 0 & 0 & 0 & 0 & 0 & 0 \\
0 & 0 & 0 & 0 & 1 & 0 & 0 & 0 \\
0 & 0 & 1 & 0 & 0 & 0 & 0 & 0 \\
0 & 0 & 0 & 0 & 0 & 0 & 1 & 0 \\
0 & 1 & 0 & 0 & 0 & 0 & 0 & 0 \\
0 & 0 & 0 & 0 & 0 & 1 & 0 & 0 \\
0 & 0 & 0 & 1 & 0 & 0 & 0 & 0 \\
0 & 0 & 0 & 0 & 0 & 0 & 0 & 1
\end{bmatrix}}_{=\hat{S}}
\begin{bmatrix} 
\psi_0 \\ \psi_0 \\ \psi_1 \\ \psi_1 \\ \psi_2 \\ \psi_2 \\ \psi_3 \\ \psi_3
\end{bmatrix}
=
\frac{1}{\sqrt{2}}
\begin{bmatrix} 
\psi_0 \\ \psi_2 \\ \psi_1 \\ \psi_3 \\ \psi_0 \\ \psi_2 \\ \psi_1 \\ \psi_3
\end{bmatrix}.
\end{align}
Note this gate only affects the velocity register qubit and the $t=2$ neighbor register qubit.

Then the coin operator, which again only affects the velocity and $t=2$ neighbor qubits, gives us the solutions for $t=2$. To construct this coin operator, recall how the coin operates on the base states [Eq. \eqref{eq:coinmatrix}]:
\begin{align}
\hat{U}_C \ket{00} &= \alpha_{00}\ket{00} + (1-\alpha_{00})\ket{01} \\
\hat{U}_C \ket{01} &= (1-\alpha_{00})\ket{00} - \alpha_{00}\ket{01} \\
\hat{U}_C \ket{10} &= (1-\alpha_{00})\ket{11} - \alpha_{00}\ket{10} \\
\hat{U}_C \ket{11} &= \alpha_{00}\ket{11} + (1-\alpha_{00})\ket{10}
\end{align}
modulo some normalization factor. For the three-qubit circuit, this becomes
\begin{align}
\hat{U}_C \ket{0k0} &= \alpha_{00}\ket{0k0} + (1-\alpha_{00})\ket{0k1} \\
\hat{U}_C \ket{0k1} &= (1-\alpha_{00})\ket{0k0} - \alpha_{00}\ket{0k1} \\
\hat{U}_C \ket{1k0} &= (1-\alpha_{00})\ket{1k1} - \alpha_{00}\ket{1k0} \\
\hat{U}_C \ket{1k1} &= \alpha_{00}\ket{1k1} + (1-\alpha_{00})\ket{1k0}
\end{align}
for $k \in \{0,1\}$. The coin operator only affects the first and third qubits corresponding to the velocity and $t=2$ neighbor qubits respectively. The second qubit denoted as $k$ is not changed.

To build the matrix, we let
\begin{widetext}
\begin{align}
\hat{U}_C
&= 
\begingroup
\setlength\arraycolsep{2pt}
\begin{bmatrix}
\bra{000}\hat{U}_C\ket{000} & \bra{000}\hat{U}_C\ket{001} & \bra{000}\hat{U}_C\ket{010} & \bra{000}\hat{U}_C\ket{011} & \bra{000}\hat{U}_C\ket{100} & \bra{000}\hat{U}_C\ket{101} & \bra{000}\hat{U}_C\ket{110} & \bra{000}\hat{U}_C\ket{111} 
\\
\bra{001}\hat{U}_C\ket{000} & \bra{001}\hat{U}_C\ket{001} & \bra{001}\hat{U}_C\ket{010} & \bra{001}\hat{U}_C\ket{011} & \bra{001}\hat{U}_C\ket{100} & \bra{001}\hat{U}_C\ket{101} & \bra{001}\hat{U}_C\ket{110} & \bra{001}\hat{U}_C\ket{111}
\\
\bra{010}\hat{U}_C\ket{000} & \bra{010}\hat{U}_C\ket{001} & \bra{010}\hat{U}_C\ket{010} & \bra{010}\hat{U}_C\ket{011} & \bra{010}\hat{U}_C\ket{100} & \bra{010}\hat{U}_C\ket{101} & \bra{010}\hat{U}_C\ket{110} & \bra{010}\hat{U}_C\ket{111} 
\\
\bra{011}\hat{U}_C\ket{000} & \bra{011}\hat{U}_C\ket{001} & \bra{011}\hat{U}_C\ket{010} & \bra{011}\hat{U}_C\ket{011} & \bra{011}\hat{U}_C\ket{100} & \bra{011}\hat{U}_C\ket{101} & \bra{011}\hat{U}_C\ket{110} & \bra{011}\hat{U}_C\ket{111} 
\\
\bra{100}\hat{U}_C\ket{000} & \bra{100}\hat{U}_C\ket{001} & \bra{100}\hat{U}_C\ket{010} & \bra{100}\hat{U}_C\ket{011} & \bra{100}\hat{U}_C\ket{100} & \bra{100}\hat{U}_C\ket{101} & \bra{100}\hat{U}_C\ket{110} & \bra{100}\hat{U}_C\ket{111} 
\\
\bra{101}\hat{U}_C\ket{000} & \bra{101}\hat{U}_C\ket{001} & \bra{101}\hat{U}_C\ket{010} & \bra{101}\hat{U}_C\ket{011} & \bra{101}\hat{U}_C\ket{100} & \bra{101}\hat{U}_C\ket{101} & \bra{101}\hat{U}_C\ket{110} & \bra{101}\hat{U}_C\ket{111}
\\
\bra{110}\hat{U}_C\ket{000} & \bra{110}\hat{U}_C\ket{001} & \bra{110}\hat{U}_C\ket{010} & \bra{110}\hat{U}_C\ket{011} & \bra{110}\hat{U}_C\ket{100} & \bra{110}\hat{U}_C\ket{101} & \bra{110}\hat{U}_C\ket{110} & \bra{110}\hat{U}_C\ket{111} 
\\
\bra{111}\hat{U}_C\ket{000} & \bra{111}\hat{U}_C\ket{001} & \bra{111}\hat{U}_C\ket{010} & \bra{111}\hat{U}_C\ket{011} & \bra{111}\hat{U}_C\ket{100} & \bra{111}\hat{U}_C\ket{101} & \bra{111}\hat{U}_C\ket{110} & \bra{111}\hat{U}_C\ket{111}
\end{bmatrix} \endgroup
\\
&= \mathcal{N}
\begin{bmatrix}
\alpha_{00} & 1-\alpha_{00} & 0 & 0 & 0 & 0 & 0 & 0 \\ 
1-\alpha_{00} & -\alpha_{00} & 0 & 0 & 0 & 0 & 0 & 0 \\
0 & 0 & \alpha_{00} & 1-\alpha_{00} & 0 & 0 & 0 & 0 \\
0 & 0 & 1-\alpha_{00} & -\alpha_{00} & 0 & 0 & 0 & 0 \\
0 & 0 & 0 & 0 & -\alpha_{00} & 1-\alpha_{00} & 0 & 0 \\
0 & 0 & 0 & 0 & 1-\alpha_{00} & \alpha_{00} & 0 & 0 \\
0 & 0 & 0 & 0 & 0 & 0 & -\alpha_{00} & 1-\alpha_{00} \\
0 & 0 & 0 & 0 & 0 & 0 & 1-\alpha_{00} & \alpha_{00}
\end{bmatrix}
\end{align}
\end{widetext}

Applying this coin then gives the statevector
\begin{align}
\hat{U}_C
\begin{bmatrix} 
\psi_0 \\ \psi_2 \\ \psi_1 \\ \psi_3 \\ \psi_0 \\ \psi_2 \\ \psi_1 \\ \psi_3
\end{bmatrix}
=
\frac{\mathcal{N}}{\sqrt{2}}
\begin{bmatrix} 
\alpha_{00}\psi_0 + (1-\alpha_{00})\psi_2 \\
(1-\alpha_{00})\psi_0 - \alpha_{00}\psi_2 \\
\alpha_{00}\psi_1 + (1-\alpha_{00})\psi_3 \\
(1-\alpha_{00})\psi_1 - \alpha_{00}\psi_3 \\
-\alpha_{00}\psi_0 + (1-\alpha_{00})\psi_2 \\
(1-\alpha_{00})\psi_0 + \alpha_{00}\psi_2 \\
-\alpha_{00}\psi_1 + (1-\alpha_{00})\psi_3 \\
(1-\alpha_{00})\psi_1 + \alpha_{00}\psi_3
\end{bmatrix}
=
\begin{bmatrix} 
\chi_0 \\ \chi_1 \\ \chi_2 \\ \chi_3 \\ \chi_4 \\ \chi_5 \\ \chi_6 \\ \chi_7
\end{bmatrix}\label{eq:chit2}
\end{align}

The classical solution [Eq. \eqref{eq:advection}] at $t=2$ is
\begin{equation}
u_{0,1}(2) = \alpha_{00} u_{0,1}(1) + (1-\alpha_{00}) u_{1,0}(1),
\end{equation}
or, using Eq. \eqref{eq:psit1},
\begin{equation}
u_{0,1}(2) = \frac{\mathcal{N}}{\sqrt{2}} ( \alpha_{00} \psi_{0,3} + (1-\alpha_{00}) \psi_{3,0}).
\end{equation}

Using Eqs. \eqref{eq:psit1} and \eqref{eq:chit2}, we can take linear combinations of $\chi_j$ elements to build $u_{0,1}(2)$:
\begin{align}
\frac{1}{2} (\chi_0-\chi_4) &= \alpha_{00} u_0(1) \\
\frac{1}{2} (\chi_1+\chi_5) &= (1-\alpha_{00}) u_0(1) \\
\frac{1}{2} (\chi_2+\chi_6) &= (1-\alpha_{00}) u_1(1) \\
\frac{1}{2} (\chi_7-\chi_3) &= \alpha_{00} u_1(1).
\end{align}

Hence,
\begin{gather}
u_0(2) = \frac{\mathcal{N}^2\mathcal{C}}{2}
\left( \frac{1}{2} (\chi_0-\chi_4) + \frac{1}{2} (\chi_2+\chi_6) \right)
\\
u_1(2) = \frac{\mathcal{N}^2\mathcal{C}}{2}
\left( \frac{1}{2} (\chi_7-\chi_3) + \frac{1}{2} (\chi_1+\chi_5) \right)
\end{gather}
where the $\mathcal{N}^2\mathcal{C}/2$ constants arise from normalizing the initial ``velocity'' register statevector ($\mathcal{C}$), normalizing two ``neighbor'' registers $(1/\sqrt{2})^2$, and applying two coin operations ($\mathcal{N}^2$).

We use similar procedures when post-processing results for the next timestep, $u_{0,1}(3)$.

\section*{References}

% \bibliography{main}

\begin{thebibliography}{66}%
\makeatletter
\providecommand \@ifxundefined [1]{%
 \@ifx{#1\undefined}
}%
\providecommand \@ifnum [1]{%
 \ifnum #1\expandafter \@firstoftwo
 \else \expandafter \@secondoftwo
 \fi
}%
\providecommand \@ifx [1]{%
 \ifx #1\expandafter \@firstoftwo
 \else \expandafter \@secondoftwo
 \fi
}%
\providecommand \natexlab [1]{#1}%
\providecommand \enquote  [1]{``#1''}%
\providecommand \bibnamefont  [1]{#1}%
\providecommand \bibfnamefont [1]{#1}%
\providecommand \citenamefont [1]{#1}%
\providecommand \href@noop [0]{\@secondoftwo}%
\providecommand \href [0]{\begingroup \@sanitize@url \@href}%
\providecommand \@href[1]{\@@startlink{#1}\@@href}%
\providecommand \@@href[1]{\endgroup#1\@@endlink}%
\providecommand \@sanitize@url [0]{\catcode `\\12\catcode `\$12\catcode `\&12\catcode `\#12\catcode `\^12\catcode `\_12\catcode `\%12\relax}%
\providecommand \@@startlink[1]{}%
\providecommand \@@endlink[0]{}%
\providecommand \url  [0]{\begingroup\@sanitize@url \@url }%
\providecommand \@url [1]{\endgroup\@href {#1}{\urlprefix }}%
\providecommand \urlprefix  [0]{URL }%
\providecommand \Eprint [0]{\href }%
\providecommand \doibase [0]{https://doi.org/}%
\providecommand \selectlanguage [0]{\@gobble}%
\providecommand \bibinfo  [0]{\@secondoftwo}%
\providecommand \bibfield  [0]{\@secondoftwo}%
\providecommand \translation [1]{[#1]}%
\providecommand \BibitemOpen [0]{}%
\providecommand \bibitemStop [0]{}%
\providecommand \bibitemNoStop [0]{.\EOS\space}%
\providecommand \EOS [0]{\spacefactor3000\relax}%
\providecommand \BibitemShut  [1]{\csname bibitem#1\endcsname}%
\let\auto@bib@innerbib\@empty
%</preamble>
\bibitem [{\citenamefont {{F. Arute et al.}}(2019)}]{Arute2019}%
  \BibitemOpen
  \bibfield  {author} {\bibinfo {author} {\bibnamefont {{F. Arute et al.}}},\ }\bibfield  {title} {\enquote {\bibinfo {title} {Quantum supremacy using a programmable superconducting processor},}\ }\href {https://doi.org/10.1038/s41586-019-1666-5} {\bibfield  {journal} {\bibinfo  {journal} {Nature}\ }\textbf {\bibinfo {volume} {574}},\ \bibinfo {pages} {505--510} (\bibinfo {year} {2019})}\BibitemShut {NoStop}%
\bibitem [{\citenamefont {{Y. Wu et al.}}(2021)}]{Wu2021}%
  \BibitemOpen
  \bibfield  {author} {\bibinfo {author} {\bibnamefont {{Y. Wu et al.}}},\ }\bibfield  {title} {\enquote {\bibinfo {title} {Strong quantum computational advantage using a superconducting quantum processor},}\ }\href {https://doi.org/10.1103/PhysRevLett.127.180501} {\bibfield  {journal} {\bibinfo  {journal} {Phys. Rev. Lett.}\ }\textbf {\bibinfo {volume} {127}},\ \bibinfo {pages} {180501} (\bibinfo {year} {2021})}\BibitemShut {NoStop}%
\bibitem [{\citenamefont {Madsen}\ \emph {et~al.}(2022)\citenamefont {Madsen}, \citenamefont {Laudenbach}, \citenamefont {Askarani}, \citenamefont {Rortais}, \citenamefont {Vincent}, \citenamefont {Bulmer}, \citenamefont {Miatto}, \citenamefont {Neuhaus}, \citenamefont {Helt}, \citenamefont {Collins}, \citenamefont {Lita}, \citenamefont {Gerrits}, \citenamefont {Nam}, \citenamefont {Vaidya}, \citenamefont {Menotti}, \citenamefont {Dhand}, \citenamefont {Vernon}, \citenamefont {Quesada},\ and\ \citenamefont {Lavoie}}]{Madsen2022}%
  \BibitemOpen
  \bibfield  {author} {\bibinfo {author} {\bibfnamefont {L.~S.}\ \bibnamefont {Madsen}}, \bibinfo {author} {\bibfnamefont {F.}~\bibnamefont {Laudenbach}}, \bibinfo {author} {\bibfnamefont {M.~F.}\ \bibnamefont {Askarani}}, \bibinfo {author} {\bibfnamefont {F.}~\bibnamefont {Rortais}}, \bibinfo {author} {\bibfnamefont {T.}~\bibnamefont {Vincent}}, \bibinfo {author} {\bibfnamefont {J.~F.~F.}\ \bibnamefont {Bulmer}}, \bibinfo {author} {\bibfnamefont {F.~M.}\ \bibnamefont {Miatto}}, \bibinfo {author} {\bibfnamefont {L.}~\bibnamefont {Neuhaus}}, \bibinfo {author} {\bibfnamefont {L.~G.}\ \bibnamefont {Helt}}, \bibinfo {author} {\bibfnamefont {M.~J.}\ \bibnamefont {Collins}}, \bibinfo {author} {\bibfnamefont {A.~E.}\ \bibnamefont {Lita}}, \bibinfo {author} {\bibfnamefont {T.}~\bibnamefont {Gerrits}}, \bibinfo {author} {\bibfnamefont {S.~W.}\ \bibnamefont {Nam}}, \bibinfo {author} {\bibfnamefont {V.~D.}\ \bibnamefont {Vaidya}}, \bibinfo {author} {\bibfnamefont {M.}~\bibnamefont {Menotti}}, \bibinfo {author}
  {\bibfnamefont {I.}~\bibnamefont {Dhand}}, \bibinfo {author} {\bibfnamefont {Z.}~\bibnamefont {Vernon}}, \bibinfo {author} {\bibfnamefont {N.}~\bibnamefont {Quesada}},\ and\ \bibinfo {author} {\bibfnamefont {J.}~\bibnamefont {Lavoie}},\ }\bibfield  {title} {\enquote {\bibinfo {title} {Quantum computational advantage with a programmable photonic processor},}\ }\href {https://doi.org/10.1038/s41586-022-04725-x} {\bibfield  {journal} {\bibinfo  {journal} {Nature}\ }\textbf {\bibinfo {volume} {606}},\ \bibinfo {pages} {75--81} (\bibinfo {year} {2022})}\BibitemShut {NoStop}%
\bibitem [{\citenamefont {Li}\ \emph {et~al.}(2025)\citenamefont {Li}, \citenamefont {Yin}, \citenamefont {Wiebe}, \citenamefont {Chun}, \citenamefont {Schenter}, \citenamefont {Cheung},\ and\ \citenamefont {M{\"u}lmenst{\"a}dt}}]{Li2025}%
  \BibitemOpen
  \bibfield  {author} {\bibinfo {author} {\bibfnamefont {X.}~\bibnamefont {Li}}, \bibinfo {author} {\bibfnamefont {X.}~\bibnamefont {Yin}}, \bibinfo {author} {\bibfnamefont {N.}~\bibnamefont {Wiebe}}, \bibinfo {author} {\bibfnamefont {J.}~\bibnamefont {Chun}}, \bibinfo {author} {\bibfnamefont {G.~K.}\ \bibnamefont {Schenter}}, \bibinfo {author} {\bibfnamefont {M.~S.}\ \bibnamefont {Cheung}},\ and\ \bibinfo {author} {\bibfnamefont {J.}~\bibnamefont {M{\"u}lmenst{\"a}dt}},\ }\bibfield  {title} {\enquote {\bibinfo {title} {Potential quantum advantage for simulation of fluid dynamics},}\ }\href {https://doi.org/10.1103/PhysRevResearch.7.013036} {\bibfield  {journal} {\bibinfo  {journal} {Phys. Rev. Res.}\ }\textbf {\bibinfo {volume} {7}},\ \bibinfo {pages} {013036} (\bibinfo {year} {2025})}\BibitemShut {NoStop}%
\bibitem [{\citenamefont {Gaitan}(2021)}]{Gaitan2021}%
  \BibitemOpen
  \bibfield  {author} {\bibinfo {author} {\bibfnamefont {F.}~\bibnamefont {Gaitan}},\ }\bibfield  {title} {\enquote {\bibinfo {title} {Finding solutions of the {Navier-Stokes} equations through quantum computing-recent progress, a generalization, and next steps forward},}\ }\href {https://doi.org/10.1002/qute.202100055} {\bibfield  {journal} {\bibinfo  {journal} {Adv. Quantum Technol.}\ }\textbf {\bibinfo {volume} {4}},\ \bibinfo {pages} {2100055} (\bibinfo {year} {2021})}\BibitemShut {NoStop}%
\bibitem [{\citenamefont {Dalzell}\ \emph {et~al.}(2025)\citenamefont {Dalzell}, \citenamefont {McArdle}, \citenamefont {Berta}, \citenamefont {Bienias}, \citenamefont {Chen}, \citenamefont {Gily{\'e}n}, \citenamefont {Hann}, \citenamefont {Kastoryano}, \citenamefont {Khabiboulline}, \citenamefont {Kubica}, \citenamefont {Salton}, \citenamefont {Wang},\ and\ \citenamefont {Brand{\~a}o}}]{Dalzell2025}%
  \BibitemOpen
  \bibfield  {author} {\bibinfo {author} {\bibfnamefont {A.~M.}\ \bibnamefont {Dalzell}}, \bibinfo {author} {\bibfnamefont {S.}~\bibnamefont {McArdle}}, \bibinfo {author} {\bibfnamefont {M.}~\bibnamefont {Berta}}, \bibinfo {author} {\bibfnamefont {P.}~\bibnamefont {Bienias}}, \bibinfo {author} {\bibfnamefont {C.-F.}\ \bibnamefont {Chen}}, \bibinfo {author} {\bibfnamefont {A.}~\bibnamefont {Gily{\'e}n}}, \bibinfo {author} {\bibfnamefont {C.~T.}\ \bibnamefont {Hann}}, \bibinfo {author} {\bibfnamefont {M.~J.}\ \bibnamefont {Kastoryano}}, \bibinfo {author} {\bibfnamefont {E.~T.}\ \bibnamefont {Khabiboulline}}, \bibinfo {author} {\bibfnamefont {A.}~\bibnamefont {Kubica}}, \bibinfo {author} {\bibfnamefont {G.}~\bibnamefont {Salton}}, \bibinfo {author} {\bibfnamefont {S.}~\bibnamefont {Wang}},\ and\ \bibinfo {author} {\bibfnamefont {F.~G. S.~L.}\ \bibnamefont {Brand{\~a}o}},\ }\href@noop {} {\emph {\bibinfo {title} {Quantum Algorithms: A Survey of Applications and End-to-End Complexities}}}\ (\bibinfo  {publisher}
  {Cambridge University Press},\ \bibinfo {address} {Cambridge},\ \bibinfo {year} {2025})\BibitemShut {NoStop}%
\bibitem [{\citenamefont {Bharadwaj}\ and\ \citenamefont {Sreenivasan}(2023{\natexlab{a}})}]{Bharadwaj2023}%
  \BibitemOpen
  \bibfield  {author} {\bibinfo {author} {\bibfnamefont {S.~S.}\ \bibnamefont {Bharadwaj}}\ and\ \bibinfo {author} {\bibfnamefont {K.~R.}\ \bibnamefont {Sreenivasan}},\ }\bibfield  {title} {\enquote {\bibinfo {title} {Hybrid quantum algorithms for flow problems},}\ }\href {https://doi.org/10.1073/pnas.2311014120} {\bibfield  {journal} {\bibinfo  {journal} {Proc. Natl. Acad. Sci.}\ }\textbf {\bibinfo {volume} {120}},\ \bibinfo {pages} {e2311014120} (\bibinfo {year} {2023}{\natexlab{a}})}\BibitemShut {NoStop}%
\bibitem [{\citenamefont {Jaksch}\ \emph {et~al.}(2023)\citenamefont {Jaksch}, \citenamefont {Givi}, \citenamefont {Daley},\ and\ \citenamefont {Rung}}]{Jaksch2023}%
  \BibitemOpen
  \bibfield  {author} {\bibinfo {author} {\bibfnamefont {D.}~\bibnamefont {Jaksch}}, \bibinfo {author} {\bibfnamefont {P.}~\bibnamefont {Givi}}, \bibinfo {author} {\bibfnamefont {A.~J.}\ \bibnamefont {Daley}},\ and\ \bibinfo {author} {\bibfnamefont {T.}~\bibnamefont {Rung}},\ }\bibfield  {title} {\enquote {\bibinfo {title} {Variational quantum algorithms for computational fluid dynamics},}\ }\href {https://doi.org/10.2514/1.J062426} {\bibfield  {journal} {\bibinfo  {journal} {AIAA J.}\ }\textbf {\bibinfo {volume} {61}},\ \bibinfo {pages} {1885--1894} (\bibinfo {year} {2023})}\BibitemShut {NoStop}%
\bibitem [{\citenamefont {Succi}\ \emph {et~al.}(2024)\citenamefont {Succi}, \citenamefont {Itani}, \citenamefont {Sanavio}, \citenamefont {Sreenivasan},\ and\ \citenamefont {Steijl}}]{Succi2024}%
  \BibitemOpen
  \bibfield  {author} {\bibinfo {author} {\bibfnamefont {S.}~\bibnamefont {Succi}}, \bibinfo {author} {\bibfnamefont {W.}~\bibnamefont {Itani}}, \bibinfo {author} {\bibfnamefont {C.}~\bibnamefont {Sanavio}}, \bibinfo {author} {\bibfnamefont {K.~R.}\ \bibnamefont {Sreenivasan}},\ and\ \bibinfo {author} {\bibfnamefont {R.}~\bibnamefont {Steijl}},\ }\bibfield  {title} {\enquote {\bibinfo {title} {Ensemble fluid simulations on quantum computers},}\ }\href {https://doi.org/10.1016/j.compfluid.2023.106148} {\bibfield  {journal} {\bibinfo  {journal} {Comput. Fluids}\ }\textbf {\bibinfo {volume} {270}},\ \bibinfo {pages} {106148} (\bibinfo {year} {2024})}\BibitemShut {NoStop}%
\bibitem [{\citenamefont {Harrow}, \citenamefont {Hassidim},\ and\ \citenamefont {Lloyd}(2009)}]{Harrow2009}%
  \BibitemOpen
  \bibfield  {author} {\bibinfo {author} {\bibfnamefont {A.~W.}\ \bibnamefont {Harrow}}, \bibinfo {author} {\bibfnamefont {A.}~\bibnamefont {Hassidim}},\ and\ \bibinfo {author} {\bibfnamefont {S.}~\bibnamefont {Lloyd}},\ }\bibfield  {title} {\enquote {\bibinfo {title} {Quantum algorithm for linear systems of equations},}\ }\href {https://doi.org/10.1103/PhysRevLett.103.150502} {\bibfield  {journal} {\bibinfo  {journal} {Phys. Rev. Lett.}\ }\textbf {\bibinfo {volume} {103}},\ \bibinfo {pages} {150502} (\bibinfo {year} {2009})}\BibitemShut {NoStop}%
\bibitem [{\citenamefont {Oz}\ \emph {et~al.}(2021)\citenamefont {Oz}, \citenamefont {Vuppala}, \citenamefont {Kara},\ and\ \citenamefont {Gaitan}}]{Oz2021}%
  \BibitemOpen
  \bibfield  {author} {\bibinfo {author} {\bibfnamefont {F.}~\bibnamefont {Oz}}, \bibinfo {author} {\bibfnamefont {R.~K. S.~S.}\ \bibnamefont {Vuppala}}, \bibinfo {author} {\bibfnamefont {K.}~\bibnamefont {Kara}},\ and\ \bibinfo {author} {\bibfnamefont {F.}~\bibnamefont {Gaitan}},\ }\bibfield  {title} {\enquote {\bibinfo {title} {Solving {Burgers'} equation with quantum computing},}\ }\href {https://doi.org/10.1007/s11128-021-03391-8} {\bibfield  {journal} {\bibinfo  {journal} {Quantum Inf. Process.}\ }\textbf {\bibinfo {volume} {21}},\ \bibinfo {pages} {30} (\bibinfo {year} {2021})}\BibitemShut {NoStop}%
\bibitem [{\citenamefont {Lapworth}(2022)}]{Lapworth2022}%
  \BibitemOpen
  \bibfield  {author} {\bibinfo {author} {\bibfnamefont {L.}~\bibnamefont {Lapworth}},\ }\href@noop {} {\enquote {\bibinfo {title} {A hybrid quantum-classical {CFD} methodology with benchmark {HHL} solutions},}\ } (\bibinfo {year} {2022}),\ \bibinfo {note} {\href{https://arxiv.org/abs/2206.00419}{arXiv:2206.00419}}\BibitemShut {NoStop}%
\bibitem [{\citenamefont {Preskill}(2018)}]{Preskill2018}%
  \BibitemOpen
  \bibfield  {author} {\bibinfo {author} {\bibfnamefont {J.}~\bibnamefont {Preskill}},\ }\bibfield  {title} {\enquote {\bibinfo {title} {Quantum computing in the {NISQ} era and beyond},}\ }\href {https://doi.org/10.22331/q-2018-08-06-79} {\bibfield  {journal} {\bibinfo  {journal} {Quantum}\ }\textbf {\bibinfo {volume} {2}},\ \bibinfo {pages} {79} (\bibinfo {year} {2018})}\BibitemShut {NoStop}%
\bibitem [{\citenamefont {Bharti}\ \emph {et~al.}(2022)\citenamefont {Bharti}, \citenamefont {Cervera-Lierta}, \citenamefont {Kyaw}, \citenamefont {Haug}, \citenamefont {Alperin-Lea}, \citenamefont {Anand}, \citenamefont {Degroote}, \citenamefont {Heimonen}, \citenamefont {Kottmann}, \citenamefont {Menke}, \citenamefont {Mok}, \citenamefont {Sim}, \citenamefont {Kwek},\ and\ \citenamefont {Aspuru-Guzik}}]{Bharti2022}%
  \BibitemOpen
  \bibfield  {author} {\bibinfo {author} {\bibfnamefont {K.}~\bibnamefont {Bharti}}, \bibinfo {author} {\bibfnamefont {A.}~\bibnamefont {Cervera-Lierta}}, \bibinfo {author} {\bibfnamefont {T.~H.}\ \bibnamefont {Kyaw}}, \bibinfo {author} {\bibfnamefont {T.}~\bibnamefont {Haug}}, \bibinfo {author} {\bibfnamefont {S.}~\bibnamefont {Alperin-Lea}}, \bibinfo {author} {\bibfnamefont {A.}~\bibnamefont {Anand}}, \bibinfo {author} {\bibfnamefont {M.}~\bibnamefont {Degroote}}, \bibinfo {author} {\bibfnamefont {H.}~\bibnamefont {Heimonen}}, \bibinfo {author} {\bibfnamefont {J.~S.}\ \bibnamefont {Kottmann}}, \bibinfo {author} {\bibfnamefont {T.}~\bibnamefont {Menke}}, \bibinfo {author} {\bibfnamefont {W.-K.}\ \bibnamefont {Mok}}, \bibinfo {author} {\bibfnamefont {S.}~\bibnamefont {Sim}}, \bibinfo {author} {\bibfnamefont {L.-C.}\ \bibnamefont {Kwek}},\ and\ \bibinfo {author} {\bibfnamefont {A.}~\bibnamefont {Aspuru-Guzik}},\ }\bibfield  {title} {\enquote {\bibinfo {title} {Noisy intermediate-scale quantum algorithms},}\
  }\href {https://doi.org/10.1103/RevModPhys.94.015004} {\bibfield  {journal} {\bibinfo  {journal} {Rev. Mod. Phys.}\ }\textbf {\bibinfo {volume} {94}},\ \bibinfo {pages} {015004} (\bibinfo {year} {2022})}\BibitemShut {NoStop}%
\bibitem [{\citenamefont {Succi}\ \emph {et~al.}(2023)\citenamefont {Succi}, \citenamefont {Itani}, \citenamefont {Sreenivasan},\ and\ \citenamefont {Steijl}}]{Succi2023}%
  \BibitemOpen
  \bibfield  {author} {\bibinfo {author} {\bibfnamefont {S.}~\bibnamefont {Succi}}, \bibinfo {author} {\bibfnamefont {W.}~\bibnamefont {Itani}}, \bibinfo {author} {\bibfnamefont {K.~R.}\ \bibnamefont {Sreenivasan}},\ and\ \bibinfo {author} {\bibfnamefont {R.}~\bibnamefont {Steijl}},\ }\bibfield  {title} {\enquote {\bibinfo {title} {Quantum computing for fluids: Where do we stand?}}\ }\href {https://doi.org/10.1209/0295-5075/acfdc7} {\bibfield  {journal} {\bibinfo  {journal} {EPL}\ }\textbf {\bibinfo {volume} {144}},\ \bibinfo {pages} {10001} (\bibinfo {year} {2023})}\BibitemShut {NoStop}%
\bibitem [{\citenamefont {Aaronson}(2015)}]{Aaronson2015}%
  \BibitemOpen
  \bibfield  {author} {\bibinfo {author} {\bibfnamefont {S.}~\bibnamefont {Aaronson}},\ }\bibfield  {title} {\enquote {\bibinfo {title} {Read the fine print},}\ }\href {https://doi.org/10.1038/nphys3272} {\bibfield  {journal} {\bibinfo  {journal} {Nat. Phys.}\ }\textbf {\bibinfo {volume} {11}},\ \bibinfo {pages} {291--293} (\bibinfo {year} {2015})}\BibitemShut {NoStop}%
\bibitem [{\citenamefont {Kim}\ \emph {et~al.}(2023)\citenamefont {Kim}, \citenamefont {Eddins}, \citenamefont {Anand}, \citenamefont {Wei}, \citenamefont {van~den Berg}, \citenamefont {Rosenblatt}, \citenamefont {Nayfeh}, \citenamefont {Wu}, \citenamefont {Zaletel}, \citenamefont {Temme},\ and\ \citenamefont {Kandala}}]{Kim2023}%
  \BibitemOpen
  \bibfield  {author} {\bibinfo {author} {\bibfnamefont {Y.}~\bibnamefont {Kim}}, \bibinfo {author} {\bibfnamefont {A.}~\bibnamefont {Eddins}}, \bibinfo {author} {\bibfnamefont {S.}~\bibnamefont {Anand}}, \bibinfo {author} {\bibfnamefont {K.~X.}\ \bibnamefont {Wei}}, \bibinfo {author} {\bibfnamefont {E.}~\bibnamefont {van~den Berg}}, \bibinfo {author} {\bibfnamefont {S.}~\bibnamefont {Rosenblatt}}, \bibinfo {author} {\bibfnamefont {H.}~\bibnamefont {Nayfeh}}, \bibinfo {author} {\bibfnamefont {Y.}~\bibnamefont {Wu}}, \bibinfo {author} {\bibfnamefont {M.}~\bibnamefont {Zaletel}}, \bibinfo {author} {\bibfnamefont {K.}~\bibnamefont {Temme}},\ and\ \bibinfo {author} {\bibfnamefont {A.}~\bibnamefont {Kandala}},\ }\bibfield  {title} {\enquote {\bibinfo {title} {Evidence for the utility of quantum computing before fault tolerance},}\ }\href {https://doi.org/10.1038/s41586-023-06096-3} {\bibfield  {journal} {\bibinfo  {journal} {Nature}\ }\textbf {\bibinfo {volume} {618}},\ \bibinfo {pages} {500--505} (\bibinfo {year}
  {2023})}\BibitemShut {NoStop}%
\bibitem [{\citenamefont {Georgescu}, \citenamefont {Ashhab},\ and\ \citenamefont {Nori}(2014)}]{Georgescu2014}%
  \BibitemOpen
  \bibfield  {author} {\bibinfo {author} {\bibfnamefont {I.~M.}\ \bibnamefont {Georgescu}}, \bibinfo {author} {\bibfnamefont {S.}~\bibnamefont {Ashhab}},\ and\ \bibinfo {author} {\bibfnamefont {F.}~\bibnamefont {Nori}},\ }\bibfield  {title} {\enquote {\bibinfo {title} {Quantum simulation},}\ }\href {https://doi.org/10.1103/RevModPhys.86.153} {\bibfield  {journal} {\bibinfo  {journal} {Rev. Mod. Phys.}\ }\textbf {\bibinfo {volume} {86}},\ \bibinfo {pages} {153--185} (\bibinfo {year} {2014})}\BibitemShut {NoStop}%
\bibitem [{\citenamefont {Succi}, \citenamefont {Fillion-Gourdeau},\ and\ \citenamefont {Palpacelli}(2015)}]{Succi2015}%
  \BibitemOpen
  \bibfield  {author} {\bibinfo {author} {\bibfnamefont {S.}~\bibnamefont {Succi}}, \bibinfo {author} {\bibfnamefont {F.}~\bibnamefont {Fillion-Gourdeau}},\ and\ \bibinfo {author} {\bibfnamefont {S.}~\bibnamefont {Palpacelli}},\ }\bibfield  {title} {\enquote {\bibinfo {title} {Quantum lattice {Boltzmann} is a quantum walk},}\ }\href {https://doi.org/10.1140/epjqt/s40507-015-0025-1} {\bibfield  {journal} {\bibinfo  {journal} {EPJ Quantum Technol.}\ }\textbf {\bibinfo {volume} {2}},\ \bibinfo {pages} {12} (\bibinfo {year} {2015})}\BibitemShut {NoStop}%
\bibitem [{\citenamefont {Itani}, \citenamefont {Sreenivasan},\ and\ \citenamefont {Succi}(2024)}]{Itani2024}%
  \BibitemOpen
  \bibfield  {author} {\bibinfo {author} {\bibfnamefont {W.}~\bibnamefont {Itani}}, \bibinfo {author} {\bibfnamefont {K.~B.}\ \bibnamefont {Sreenivasan}},\ and\ \bibinfo {author} {\bibfnamefont {S.}~\bibnamefont {Succi}},\ }\bibfield  {title} {\enquote {\bibinfo {title} {Quantum algorithm for lattice {Boltzmann} ({QALB}) simulation of incompressible fluids with a nonlinear collision term},}\ }\href {https://doi.org/10.1063/5.0176569} {\bibfield  {journal} {\bibinfo  {journal} {Phys. Fluids}\ }\textbf {\bibinfo {volume} {36}},\ \bibinfo {pages} {017112} (\bibinfo {year} {2024})}\BibitemShut {NoStop}%
\bibitem [{\citenamefont {Sanavio}\ and\ \citenamefont {Succi}(2024)}]{Sanavio2024}%
  \BibitemOpen
  \bibfield  {author} {\bibinfo {author} {\bibfnamefont {C.}~\bibnamefont {Sanavio}}\ and\ \bibinfo {author} {\bibfnamefont {S.}~\bibnamefont {Succi}},\ }\bibfield  {title} {\enquote {\bibinfo {title} {Lattice {Boltzmann-Carleman} quantum algorithm and circuit for fluid flows at moderate {Reynolds} number},}\ }\href {https://doi.org/10.1116/5.0195549} {\bibfield  {journal} {\bibinfo  {journal} {AVS Quantum Sci.}\ }\textbf {\bibinfo {volume} {6}},\ \bibinfo {pages} {023802} (\bibinfo {year} {2024})}\BibitemShut {NoStop}%
\bibitem [{\citenamefont {Budinski}(2021)}]{Budinski2021}%
  \BibitemOpen
  \bibfield  {author} {\bibinfo {author} {\bibfnamefont {L.}~\bibnamefont {Budinski}},\ }\bibfield  {title} {\enquote {\bibinfo {title} {Quantum algorithm for the advection-diffusion equation simulated with the lattice {Boltzmann} method},}\ }\href {https://doi.org/10.1007/s11128-021-02996-3} {\bibfield  {journal} {\bibinfo  {journal} {Quantum Inf. Process.}\ }\textbf {\bibinfo {volume} {20}},\ \bibinfo {pages} {57} (\bibinfo {year} {2021})}\BibitemShut {NoStop}%
\bibitem [{\citenamefont {Aharonov}, \citenamefont {Davidovich},\ and\ \citenamefont {Zagury}(1992)}]{Aharonov1992}%
  \BibitemOpen
  \bibfield  {author} {\bibinfo {author} {\bibfnamefont {Y.}~\bibnamefont {Aharonov}}, \bibinfo {author} {\bibfnamefont {L.}~\bibnamefont {Davidovich}},\ and\ \bibinfo {author} {\bibfnamefont {N.}~\bibnamefont {Zagury}},\ }\bibfield  {title} {\enquote {\bibinfo {title} {Quantum random walks},}\ }\href {https://doi.org/10.1103/PhysRevA.48.1687} {\bibfield  {journal} {\bibinfo  {journal} {Phys. Rev. A}\ }\textbf {\bibinfo {volume} {48}},\ \bibinfo {pages} {1687--1690} (\bibinfo {year} {1992})}\BibitemShut {NoStop}%
\bibitem [{\citenamefont {Farhi}\ and\ \citenamefont {Gutmann}(1998)}]{Farhi1998}%
  \BibitemOpen
  \bibfield  {author} {\bibinfo {author} {\bibfnamefont {E.}~\bibnamefont {Farhi}}\ and\ \bibinfo {author} {\bibfnamefont {S.}~\bibnamefont {Gutmann}},\ }\bibfield  {title} {\enquote {\bibinfo {title} {Quantum computation and decision trees},}\ }\href {https://doi.org/10.1103/PhysRevA.58.915} {\bibfield  {journal} {\bibinfo  {journal} {Phys. Rev. A}\ }\textbf {\bibinfo {volume} {58}},\ \bibinfo {pages} {915--928} (\bibinfo {year} {1998})}\BibitemShut {NoStop}%
\bibitem [{\citenamefont {Childs}\ \emph {et~al.}(2003)\citenamefont {Childs}, \citenamefont {Cleve}, \citenamefont {Deotto}, \citenamefont {Farhi}, \citenamefont {Gutmann},\ and\ \citenamefont {Spielman}}]{Childs2003}%
  \BibitemOpen
  \bibfield  {author} {\bibinfo {author} {\bibfnamefont {A.~M.}\ \bibnamefont {Childs}}, \bibinfo {author} {\bibfnamefont {R.}~\bibnamefont {Cleve}}, \bibinfo {author} {\bibfnamefont {E.}~\bibnamefont {Deotto}}, \bibinfo {author} {\bibfnamefont {E.}~\bibnamefont {Farhi}}, \bibinfo {author} {\bibfnamefont {S.}~\bibnamefont {Gutmann}},\ and\ \bibinfo {author} {\bibfnamefont {D.~A.}\ \bibnamefont {Spielman}},\ }\bibfield  {title} {\enquote {\bibinfo {title} {Exponential algorithmic speedup by a quantum walk},}\ }in\ \href {https://doi.org/10.1145/780542.780552} {\emph {\bibinfo {booktitle} {Proc. 35th ACM STOC}}}\ (\bibinfo  {publisher} {ACM},\ \bibinfo {address} {New York, NY},\ \bibinfo {year} {2003})\ pp.\ \bibinfo {pages} {59--68}\BibitemShut {NoStop}%
\bibitem [{\citenamefont {Lucy}(1977)}]{Lucy1977}%
  \BibitemOpen
  \bibfield  {author} {\bibinfo {author} {\bibfnamefont {L.~B.}\ \bibnamefont {Lucy}},\ }\bibfield  {title} {\enquote {\bibinfo {title} {A numerical approach to the testing of the fission hypothesis},}\ }\href {https://doi.org/10.1086/112164} {\bibfield  {journal} {\bibinfo  {journal} {Astron. J.}\ }\textbf {\bibinfo {volume} {82}},\ \bibinfo {pages} {1013--1024} (\bibinfo {year} {1977})}\BibitemShut {NoStop}%
\bibitem [{\citenamefont {Gingold}\ and\ \citenamefont {Monaghan}(1977)}]{Gingold1977}%
  \BibitemOpen
  \bibfield  {author} {\bibinfo {author} {\bibfnamefont {R.~A.}\ \bibnamefont {Gingold}}\ and\ \bibinfo {author} {\bibfnamefont {J.~J.}\ \bibnamefont {Monaghan}},\ }\bibfield  {title} {\enquote {\bibinfo {title} {Smoothed particle hydrodynamics: theory and application to non-spherical stars},}\ }\href {https://doi.org/10.1093/mnras/181.3.375} {\bibfield  {journal} {\bibinfo  {journal} {Mon. Not. R. Astron. Soc.}\ }\textbf {\bibinfo {volume} {181}},\ \bibinfo {pages} {375--389} (\bibinfo {year} {1977})}\BibitemShut {NoStop}%
\bibitem [{\citenamefont {Au-Yeung}\ \emph {et~al.}(2024)\citenamefont {Au-Yeung}, \citenamefont {Williams}, \citenamefont {Kendon},\ and\ \citenamefont {Lind}}]{AuYeung2024}%
  \BibitemOpen
  \bibfield  {author} {\bibinfo {author} {\bibfnamefont {R.}~\bibnamefont {Au-Yeung}}, \bibinfo {author} {\bibfnamefont {A.~J.}\ \bibnamefont {Williams}}, \bibinfo {author} {\bibfnamefont {V.~M.}\ \bibnamefont {Kendon}},\ and\ \bibinfo {author} {\bibfnamefont {S.~J.}\ \bibnamefont {Lind}},\ }\bibfield  {title} {\enquote {\bibinfo {title} {Quantum algorithm for smoothed particle hydrodynamics},}\ }\href {https://doi.org/10.1016/j.cpc.2023.108909} {\bibfield  {journal} {\bibinfo  {journal} {Comput. Phys. Commun.}\ }\textbf {\bibinfo {volume} {294}},\ \bibinfo {pages} {108909} (\bibinfo {year} {2024})}\BibitemShut {NoStop}%
\bibitem [{\citenamefont {Bharadwaj}\ and\ \citenamefont {Sreenivasan}(2023{\natexlab{b}})}]{Bharadwaj2023r}%
  \BibitemOpen
  \bibfield  {author} {\bibinfo {author} {\bibfnamefont {S.~S.}\ \bibnamefont {Bharadwaj}}\ and\ \bibinfo {author} {\bibfnamefont {K.~R.}\ \bibnamefont {Sreenivasan}},\ }\bibfield  {title} {\enquote {\bibinfo {title} {An introduction to algorithms in quantum computation of fluid dynamics},}\ }in\ \href@noop {} {\emph {\bibinfo {booktitle} {Introduction to Quantum Computing in Fluid Dynamics}}}\ (\bibinfo  {publisher} {NATO Science \& Technology Organization (STO)},\ \bibinfo {address} {STO Educational Notes STO-EN-AVT-377},\ \bibinfo {year} {2023})\BibitemShut {NoStop}%
\bibitem [{\citenamefont {Bharadwaj}\ and\ \citenamefont {Sreenivasan}(2020)}]{Bharadwaj2020}%
  \BibitemOpen
  \bibfield  {author} {\bibinfo {author} {\bibfnamefont {S.~S.}\ \bibnamefont {Bharadwaj}}\ and\ \bibinfo {author} {\bibfnamefont {K.~R.}\ \bibnamefont {Sreenivasan}},\ }\bibfield  {title} {\enquote {\bibinfo {title} {Quantum computation of fluid dynamics},}\ }\href {https://doi.org/10.29195/iascs.03.01.0015} {\bibfield  {journal} {\bibinfo  {journal} {Indian Acad. Sci. Conf. Ser.}\ }\textbf {\bibinfo {volume} {3}},\ \bibinfo {pages} {77--96} (\bibinfo {year} {2020})}\BibitemShut {NoStop}%
\bibitem [{\citenamefont {Givi}\ \emph {et~al.}(2020)\citenamefont {Givi}, \citenamefont {Daley}, \citenamefont {Mavriplis},\ and\ \citenamefont {Malik}}]{Givi2020}%
  \BibitemOpen
  \bibfield  {author} {\bibinfo {author} {\bibfnamefont {P.}~\bibnamefont {Givi}}, \bibinfo {author} {\bibfnamefont {A.~J.}\ \bibnamefont {Daley}}, \bibinfo {author} {\bibfnamefont {D.}~\bibnamefont {Mavriplis}},\ and\ \bibinfo {author} {\bibfnamefont {M.}~\bibnamefont {Malik}},\ }\bibfield  {title} {\enquote {\bibinfo {title} {Quantum speedup for aeroscience and engineering},}\ }\href {https://doi.org/10.2514/1.J059183} {\bibfield  {journal} {\bibinfo  {journal} {AIAA J.}\ }\textbf {\bibinfo {volume} {58}},\ \bibinfo {pages} {3715} (\bibinfo {year} {2020})}\BibitemShut {NoStop}%
\bibitem [{\citenamefont {Nielsen}\ and\ \citenamefont {Chuang}(2010)}]{Nielsen2010}%
  \BibitemOpen
  \bibfield  {author} {\bibinfo {author} {\bibfnamefont {M.~A.}\ \bibnamefont {Nielsen}}\ and\ \bibinfo {author} {\bibfnamefont {I.~L.}\ \bibnamefont {Chuang}},\ }\href {https://doi.org/10.1017/CBO9780511976667} {\emph {\bibinfo {title} {Quantum Computation and Quantum Information: 10th Anniversary Edition}}}\ (\bibinfo  {publisher} {Cambridge University Press},\ \bibinfo {year} {2010})\BibitemShut {NoStop}%
\bibitem [{\citenamefont {Monaghan}(2012)}]{Monaghan2012}%
  \BibitemOpen
  \bibfield  {author} {\bibinfo {author} {\bibfnamefont {J.~J.}\ \bibnamefont {Monaghan}},\ }\bibfield  {title} {\enquote {\bibinfo {title} {Smoothed particle hydrodynamics and its diverse applications},}\ }\href {https://doi.org/10.1146/annurev-fluid-120710-101220} {\bibfield  {journal} {\bibinfo  {journal} {Annu. Rev. Fluid Mech.}\ }\textbf {\bibinfo {volume} {44}},\ \bibinfo {pages} {323--346} (\bibinfo {year} {2012})}\BibitemShut {NoStop}%
\bibitem [{\citenamefont {Koschier}\ \emph {et~al.}(2022)\citenamefont {Koschier}, \citenamefont {Bender}, \citenamefont {Solenthaler},\ and\ \citenamefont {Teschner}}]{Koschier2022}%
  \BibitemOpen
  \bibfield  {author} {\bibinfo {author} {\bibfnamefont {D.}~\bibnamefont {Koschier}}, \bibinfo {author} {\bibfnamefont {J.}~\bibnamefont {Bender}}, \bibinfo {author} {\bibfnamefont {B.}~\bibnamefont {Solenthaler}},\ and\ \bibinfo {author} {\bibfnamefont {M.}~\bibnamefont {Teschner}},\ }\bibfield  {title} {\enquote {\bibinfo {title} {A survey on {SPH} methods in computer graphics},}\ }\href {https://doi.org/10.1111/cgf.14508} {\bibfield  {journal} {\bibinfo  {journal} {Comput. Graph. Forum}\ }\textbf {\bibinfo {volume} {41}},\ \bibinfo {pages} {737--760} (\bibinfo {year} {2022})}\BibitemShut {NoStop}%
\bibitem [{\citenamefont {Monaghan}(2005)}]{Monaghan2005}%
  \BibitemOpen
  \bibfield  {author} {\bibinfo {author} {\bibfnamefont {J.~J.}\ \bibnamefont {Monaghan}},\ }\bibfield  {title} {\enquote {\bibinfo {title} {Smoothed particle hydrodynamics},}\ }\href {https://doi.org/10.1088/0034-4885/68/8/r01} {\bibfield  {journal} {\bibinfo  {journal} {Rep. Prog. Phys.}\ }\textbf {\bibinfo {volume} {68}},\ \bibinfo {pages} {1703--1759} (\bibinfo {year} {2005})}\BibitemShut {NoStop}%
\bibitem [{\citenamefont {Lind}, \citenamefont {Rogers},\ and\ \citenamefont {Stansby}(2020)}]{Lind2020}%
  \BibitemOpen
  \bibfield  {author} {\bibinfo {author} {\bibfnamefont {S.~J.}\ \bibnamefont {Lind}}, \bibinfo {author} {\bibfnamefont {B.~D.}\ \bibnamefont {Rogers}},\ and\ \bibinfo {author} {\bibfnamefont {P.~K.}\ \bibnamefont {Stansby}},\ }\bibfield  {title} {\enquote {\bibinfo {title} {Review of smoothed particle hydrodynamics: towards converged {Lagrangian} flow modelling},}\ }\href {https://doi.org/10.1098/rspa.2019.0801} {\bibfield  {journal} {\bibinfo  {journal} {Proc. R. Soc. A}\ }\textbf {\bibinfo {volume} {476}},\ \bibinfo {pages} {20190801} (\bibinfo {year} {2020})}\BibitemShut {NoStop}%
\bibitem [{\citenamefont {Abaqus}(2017)}]{Abaqus}%
  \BibitemOpen
  \bibfield  {author} {\bibinfo {author} {\bibnamefont {Abaqus}},\ }\href@noop {} {\enquote {\bibinfo {title} {Smoothed particle hydrodynamics},}\ }\bibinfo {howpublished} {\url{https://abaqus-docs.mit.edu/2017/English/SIMACAEANLRefMap/simaanl-c-sphanalysis.htm}} (\bibinfo {year} {2017}),\ \bibinfo {note} {accessed: 2024-01-01}\BibitemShut {NoStop}%
\bibitem [{\citenamefont {Lind}\ and\ \citenamefont {Stansby}(2016)}]{lind2016high}%
  \BibitemOpen
  \bibfield  {author} {\bibinfo {author} {\bibfnamefont {S.~J.}\ \bibnamefont {Lind}}\ and\ \bibinfo {author} {\bibfnamefont {P.~K.}\ \bibnamefont {Stansby}},\ }\bibfield  {title} {\enquote {\bibinfo {title} {High-order {Eulerian} incompressible smoothed particle hydrodynamics with transition to {Lagrangian} free-surface motion},}\ }\href {https://doi.org/10.1016/j.jcp.2016.08.047} {\bibfield  {journal} {\bibinfo  {journal} {J. Comput. Phys.}\ }\textbf {\bibinfo {volume} {326}},\ \bibinfo {pages} {290--311} (\bibinfo {year} {2016})}\BibitemShut {NoStop}%
\bibitem [{\citenamefont {Nasar}\ \emph {et~al.}(2019)\citenamefont {Nasar}, \citenamefont {Rogers}, \citenamefont {Revell}, \citenamefont {Stansby},\ and\ \citenamefont {Lind}}]{nasar2019eulerian}%
  \BibitemOpen
  \bibfield  {author} {\bibinfo {author} {\bibfnamefont {A.~M.~A.}\ \bibnamefont {Nasar}}, \bibinfo {author} {\bibfnamefont {B.~D.}\ \bibnamefont {Rogers}}, \bibinfo {author} {\bibfnamefont {A.}~\bibnamefont {Revell}}, \bibinfo {author} {\bibfnamefont {P.~K.}\ \bibnamefont {Stansby}},\ and\ \bibinfo {author} {\bibfnamefont {S.~J.}\ \bibnamefont {Lind}},\ }\bibfield  {title} {\enquote {\bibinfo {title} {Eulerian weakly compressible smoothed particle hydrodynamics ({SPH}) with the immersed boundary method for thin slender bodies},}\ }\href {https://doi.org/10.1016/j.jfluidstructs.2018.11.005} {\bibfield  {journal} {\bibinfo  {journal} {J. Fluid. Struct.}\ }\textbf {\bibinfo {volume} {84}},\ \bibinfo {pages} {263--282} (\bibinfo {year} {2019})}\BibitemShut {NoStop}%
\bibitem [{\citenamefont {Nasar}\ \emph {et~al.}(2021)\citenamefont {Nasar}, \citenamefont {Fourtakas}, \citenamefont {Lind}, \citenamefont {King}, \citenamefont {Rogers},\ and\ \citenamefont {Stansby}}]{nasar2021high}%
  \BibitemOpen
  \bibfield  {author} {\bibinfo {author} {\bibfnamefont {A.~M.~A.}\ \bibnamefont {Nasar}}, \bibinfo {author} {\bibfnamefont {G.}~\bibnamefont {Fourtakas}}, \bibinfo {author} {\bibfnamefont {S.~J.}\ \bibnamefont {Lind}}, \bibinfo {author} {\bibfnamefont {J.~R.~C.}\ \bibnamefont {King}}, \bibinfo {author} {\bibfnamefont {B.~D.}\ \bibnamefont {Rogers}},\ and\ \bibinfo {author} {\bibfnamefont {P.~K.}\ \bibnamefont {Stansby}},\ }\bibfield  {title} {\enquote {\bibinfo {title} {High-order consistent {SPH} with the pressure projection method in {2-D} and {3-D}},}\ }\href {https://doi.org/10.1016/j.jcp.2021.110563} {\bibfield  {journal} {\bibinfo  {journal} {Journal of Computational Physics}\ }\textbf {\bibinfo {volume} {444}},\ \bibinfo {pages} {110563} (\bibinfo {year} {2021})}\BibitemShut {NoStop}%
\bibitem [{\citenamefont {Barenco}\ \emph {et~al.}(1997)\citenamefont {Barenco}, \citenamefont {Berthiaume}, \citenamefont {Deutsch}, \citenamefont {Ekert}, \citenamefont {Jozsa},\ and\ \citenamefont {Macchiavello}}]{Barenco1997}%
  \BibitemOpen
  \bibfield  {author} {\bibinfo {author} {\bibfnamefont {A.}~\bibnamefont {Barenco}}, \bibinfo {author} {\bibfnamefont {A.}~\bibnamefont {Berthiaume}}, \bibinfo {author} {\bibfnamefont {D.}~\bibnamefont {Deutsch}}, \bibinfo {author} {\bibfnamefont {A.}~\bibnamefont {Ekert}}, \bibinfo {author} {\bibfnamefont {R.}~\bibnamefont {Jozsa}},\ and\ \bibinfo {author} {\bibfnamefont {C.}~\bibnamefont {Macchiavello}},\ }\bibfield  {title} {\enquote {\bibinfo {title} {Stabilization of quantum computations by symmetrization},}\ }\href {https://doi.org/10.1137/S0097539796302452} {\bibfield  {journal} {\bibinfo  {journal} {SIAM J. Comput.}\ }\textbf {\bibinfo {volume} {26}},\ \bibinfo {pages} {1541--1557} (\bibinfo {year} {1997})}\BibitemShut {NoStop}%
\bibitem [{\citenamefont {Buhrman}\ \emph {et~al.}(2001)\citenamefont {Buhrman}, \citenamefont {Cleve}, \citenamefont {Watrous},\ and\ \citenamefont {de~Wolf}}]{Buhrman2001}%
  \BibitemOpen
  \bibfield  {author} {\bibinfo {author} {\bibfnamefont {H.}~\bibnamefont {Buhrman}}, \bibinfo {author} {\bibfnamefont {R.}~\bibnamefont {Cleve}}, \bibinfo {author} {\bibfnamefont {J.}~\bibnamefont {Watrous}},\ and\ \bibinfo {author} {\bibfnamefont {R.}~\bibnamefont {de~Wolf}},\ }\bibfield  {title} {\enquote {\bibinfo {title} {Quantum fingerprinting},}\ }\href {https://doi.org/10.1103/PhysRevLett.87.167902} {\bibfield  {journal} {\bibinfo  {journal} {Phys. Rev. Lett.}\ }\textbf {\bibinfo {volume} {87}},\ \bibinfo {pages} {167902} (\bibinfo {year} {2001})}\BibitemShut {NoStop}%
\bibitem [{\citenamefont {Long}\ and\ \citenamefont {Sun}(2001)}]{Long2001}%
  \BibitemOpen
  \bibfield  {author} {\bibinfo {author} {\bibfnamefont {G.-L.}\ \bibnamefont {Long}}\ and\ \bibinfo {author} {\bibfnamefont {Y.}~\bibnamefont {Sun}},\ }\bibfield  {title} {\enquote {\bibinfo {title} {Efficient scheme for initializing a quantum register with an arbitrary superposed state},}\ }\href {https://doi.org/10.1103/PhysRevA.64.014303} {\bibfield  {journal} {\bibinfo  {journal} {Phys. Rev. A}\ }\textbf {\bibinfo {volume} {64}},\ \bibinfo {pages} {014303} (\bibinfo {year} {2001})}\BibitemShut {NoStop}%
\bibitem [{\citenamefont {M{\"o}tt{\"o}nen}\ \emph {et~al.}(2005)\citenamefont {M{\"o}tt{\"o}nen}, \citenamefont {Vartiainen}, \citenamefont {Bergholm},\ and\ \citenamefont {Salomaa}}]{Mottonen2005}%
  \BibitemOpen
  \bibfield  {author} {\bibinfo {author} {\bibfnamefont {M.}~\bibnamefont {M{\"o}tt{\"o}nen}}, \bibinfo {author} {\bibfnamefont {J.~J.}\ \bibnamefont {Vartiainen}}, \bibinfo {author} {\bibfnamefont {V.}~\bibnamefont {Bergholm}},\ and\ \bibinfo {author} {\bibfnamefont {M.~M.}\ \bibnamefont {Salomaa}},\ }\bibfield  {title} {\enquote {\bibinfo {title} {Transformation of quantum states using uniformly controlled rotations},}\ }\href {https://doi.org/10.26421/QIC5.6-5} {\bibfield  {journal} {\bibinfo  {journal} {Quantum Inf. Comput.}\ }\textbf {\bibinfo {volume} {5}},\ \bibinfo {pages} {467--473} (\bibinfo {year} {2005})}\BibitemShut {NoStop}%
\bibitem [{\citenamefont {Fanizza}\ \emph {et~al.}(2020)\citenamefont {Fanizza}, \citenamefont {Rosati}, \citenamefont {Skotiniotis}, \citenamefont {Calsamiglia},\ and\ \citenamefont {Giovannetti}}]{Fanizza2020}%
  \BibitemOpen
  \bibfield  {author} {\bibinfo {author} {\bibfnamefont {M.}~\bibnamefont {Fanizza}}, \bibinfo {author} {\bibfnamefont {M.}~\bibnamefont {Rosati}}, \bibinfo {author} {\bibfnamefont {M.}~\bibnamefont {Skotiniotis}}, \bibinfo {author} {\bibfnamefont {J.}~\bibnamefont {Calsamiglia}},\ and\ \bibinfo {author} {\bibfnamefont {V.}~\bibnamefont {Giovannetti}},\ }\bibfield  {title} {\enquote {\bibinfo {title} {Beyond the swap test: Optimal estimation of quantum state overlap},}\ }\href {https://doi.org/10.1103/PhysRevLett.124.060503} {\bibfield  {journal} {\bibinfo  {journal} {Phys. Rev. Lett.}\ }\textbf {\bibinfo {volume} {124}},\ \bibinfo {pages} {060503} (\bibinfo {year} {2020})}\BibitemShut {NoStop}%
\bibitem [{\citenamefont {Ambainis}(2003)}]{Ambainis2003}%
  \BibitemOpen
  \bibfield  {author} {\bibinfo {author} {\bibfnamefont {A.}~\bibnamefont {Ambainis}},\ }\bibfield  {title} {\enquote {\bibinfo {title} {Quantum walks and their algorithmic applications},}\ }\href {https://doi.org/10.1142/S0219749903000383} {\bibfield  {journal} {\bibinfo  {journal} {Int. J. Quantum Inf.}\ }\textbf {\bibinfo {volume} {1}},\ \bibinfo {pages} {507--518} (\bibinfo {year} {2003})}\BibitemShut {NoStop}%
\bibitem [{\citenamefont {Venegas-Andraca}(2012)}]{VenegasAndraca2012}%
  \BibitemOpen
  \bibfield  {author} {\bibinfo {author} {\bibfnamefont {S.~E.}\ \bibnamefont {Venegas-Andraca}},\ }\bibfield  {title} {\enquote {\bibinfo {title} {Quantum walks: a comprehensive review},}\ }\href {https://doi.org/10.1007/s11128-012-0432-5} {\bibfield  {journal} {\bibinfo  {journal} {Quantum Inf. Process.}\ }\textbf {\bibinfo {volume} {11}},\ \bibinfo {pages} {1015--1106} (\bibinfo {year} {2012})}\BibitemShut {NoStop}%
\bibitem [{\citenamefont {Kadian}, \citenamefont {Garhwal},\ and\ \citenamefont {Kumar}(2021)}]{Kadian2021}%
  \BibitemOpen
  \bibfield  {author} {\bibinfo {author} {\bibfnamefont {K.}~\bibnamefont {Kadian}}, \bibinfo {author} {\bibfnamefont {S.}~\bibnamefont {Garhwal}},\ and\ \bibinfo {author} {\bibfnamefont {A.}~\bibnamefont {Kumar}},\ }\bibfield  {title} {\enquote {\bibinfo {title} {Quantum walk and its application domains: A systematic review},}\ }\href {https://doi.org/10.1016/j.cosrev.2021.100419} {\bibfield  {journal} {\bibinfo  {journal} {Comput. Sci. Rev.}\ }\textbf {\bibinfo {volume} {41}},\ \bibinfo {pages} {100419} (\bibinfo {year} {2021})}\BibitemShut {NoStop}%
\bibitem [{\citenamefont {Claudon}, \citenamefont {Piquemal},\ and\ \citenamefont {Monmarch{\'e}}(2025)}]{Claudon2025}%
  \BibitemOpen
  \bibfield  {author} {\bibinfo {author} {\bibfnamefont {B.}~\bibnamefont {Claudon}}, \bibinfo {author} {\bibfnamefont {J.-P.}\ \bibnamefont {Piquemal}},\ and\ \bibinfo {author} {\bibfnamefont {P.}~\bibnamefont {Monmarch{\'e}}},\ }\href@noop {} {\enquote {\bibinfo {title} {Quantum speedup for nonreversible {Markov} chains},}\ } (\bibinfo {year} {2025}),\ \bibinfo {note} {\href{https://arxiv.org/abs/2501.05868}{arXiv:2501.05868}}\BibitemShut {NoStop}%
\bibitem [{\citenamefont {Steijl}\ and\ \citenamefont {Barakos}(2018)}]{Steijl2018}%
  \BibitemOpen
  \bibfield  {author} {\bibinfo {author} {\bibfnamefont {R.}~\bibnamefont {Steijl}}\ and\ \bibinfo {author} {\bibfnamefont {G.~N.}\ \bibnamefont {Barakos}},\ }\bibfield  {title} {\enquote {\bibinfo {title} {Parallel evaluation of quantum algorithms for computational fluid dynamics},}\ }\href {https://doi.org/10.1016/j.compfluid.2018.03.080} {\bibfield  {journal} {\bibinfo  {journal} {Comput. Fluids}\ }\textbf {\bibinfo {volume} {173}},\ \bibinfo {pages} {22--28} (\bibinfo {year} {2018})}\BibitemShut {NoStop}%
\bibitem [{\citenamefont {Zylberman}\ \emph {et~al.}(2022)\citenamefont {Zylberman}, \citenamefont {Molfetta}, \citenamefont {Brachet}, \citenamefont {Loureiro},\ and\ \citenamefont {Debbasch}}]{Zylberman2022}%
  \BibitemOpen
  \bibfield  {author} {\bibinfo {author} {\bibfnamefont {J.}~\bibnamefont {Zylberman}}, \bibinfo {author} {\bibfnamefont {G.~D.}\ \bibnamefont {Molfetta}}, \bibinfo {author} {\bibfnamefont {M.}~\bibnamefont {Brachet}}, \bibinfo {author} {\bibfnamefont {N.~F.}\ \bibnamefont {Loureiro}},\ and\ \bibinfo {author} {\bibfnamefont {F.}~\bibnamefont {Debbasch}},\ }\bibfield  {title} {\enquote {\bibinfo {title} {Quantum simulations of hydrodynamics via the {Madelung} transformation},}\ }\href {https://doi.org/10.1103/PhysRevA.106.032408} {\bibfield  {journal} {\bibinfo  {journal} {Phys. Rev. A}\ }\textbf {\bibinfo {volume} {106}},\ \bibinfo {pages} {032408} (\bibinfo {year} {2022})}\BibitemShut {NoStop}%
\bibitem [{\citenamefont {Childs}(2009)}]{Childs2009}%
  \BibitemOpen
  \bibfield  {author} {\bibinfo {author} {\bibfnamefont {A.~M.}\ \bibnamefont {Childs}},\ }\bibfield  {title} {\enquote {\bibinfo {title} {Universal computation by quantum walk},}\ }\href {https://doi.org/10.1103/PhysRevLett.102.180501} {\bibfield  {journal} {\bibinfo  {journal} {Phys. Rev. Lett.}\ }\textbf {\bibinfo {volume} {102}},\ \bibinfo {pages} {180501} (\bibinfo {year} {2009})}\BibitemShut {NoStop}%
\bibitem [{\citenamefont {Childs}(2010)}]{Childs2010}%
  \BibitemOpen
  \bibfield  {author} {\bibinfo {author} {\bibfnamefont {A.~M.}\ \bibnamefont {Childs}},\ }\bibfield  {title} {\enquote {\bibinfo {title} {On the relationship between continuous- and discrete-time quantum walk},}\ }\href {https://doi.org/10.1007/s00220-009-0930-1} {\bibfield  {journal} {\bibinfo  {journal} {Commun. Math. Phys.}\ }\textbf {\bibinfo {volume} {294}},\ \bibinfo {pages} {581--603} (\bibinfo {year} {2010})}\BibitemShut {NoStop}%
\bibitem [{\citenamefont {Brassard}\ and\ \citenamefont {H{\o}yer}(1997)}]{Brassard1997}%
  \BibitemOpen
  \bibfield  {author} {\bibinfo {author} {\bibfnamefont {G.}~\bibnamefont {Brassard}}\ and\ \bibinfo {author} {\bibfnamefont {P.}~\bibnamefont {H{\o}yer}},\ }\bibfield  {title} {\enquote {\bibinfo {title} {An exact quantum polynomial-time algorithm for {Simon}'s problem},}\ }in\ \href {https://doi.org/10.1109/ISTCS.1997.595153} {\emph {\bibinfo {booktitle} {Proc. 5th ISTCS}}}\ (\bibinfo {year} {1997})\ pp.\ \bibinfo {pages} {12--23}\BibitemShut {NoStop}%
\bibitem [{\citenamefont {Grover}(1998)}]{Grover1998}%
  \BibitemOpen
  \bibfield  {author} {\bibinfo {author} {\bibfnamefont {L.~K.}\ \bibnamefont {Grover}},\ }\bibfield  {title} {\enquote {\bibinfo {title} {Quantum computers can search rapidly by using almost any transformation},}\ }\href {https://doi.org/10.1103/PhysRevLett.80.4329} {\bibfield  {journal} {\bibinfo  {journal} {Phys. Rev. Lett.}\ }\textbf {\bibinfo {volume} {80}},\ \bibinfo {pages} {4329} (\bibinfo {year} {1998})}\BibitemShut {NoStop}%
\bibitem [{\citenamefont {Bennett}\ \emph {et~al.}(1997)\citenamefont {Bennett}, \citenamefont {Bernstein}, \citenamefont {Brassard},\ and\ \citenamefont {Vazirani}}]{Bennett1997}%
  \BibitemOpen
  \bibfield  {author} {\bibinfo {author} {\bibfnamefont {C.~H.}\ \bibnamefont {Bennett}}, \bibinfo {author} {\bibfnamefont {E.}~\bibnamefont {Bernstein}}, \bibinfo {author} {\bibfnamefont {G.}~\bibnamefont {Brassard}},\ and\ \bibinfo {author} {\bibfnamefont {U.}~\bibnamefont {Vazirani}},\ }\bibfield  {title} {\enquote {\bibinfo {title} {Strengths and weaknesses of quantum computing},}\ }\href {https://doi.org/10.1137/S0097539796300933} {\bibfield  {journal} {\bibinfo  {journal} {SIAM J. Comput.}\ }\textbf {\bibinfo {volume} {26}},\ \bibinfo {pages} {1510--1523} (\bibinfo {year} {1997})}\BibitemShut {NoStop}%
\bibitem [{\citenamefont {Cleve}\ \emph {et~al.}(2013)\citenamefont {Cleve}, \citenamefont {van Dam}, \citenamefont {Nielsen},\ and\ \citenamefont {Tapp}}]{Cleve2013}%
  \BibitemOpen
  \bibfield  {author} {\bibinfo {author} {\bibfnamefont {R.}~\bibnamefont {Cleve}}, \bibinfo {author} {\bibfnamefont {W.}~\bibnamefont {van Dam}}, \bibinfo {author} {\bibfnamefont {M.}~\bibnamefont {Nielsen}},\ and\ \bibinfo {author} {\bibfnamefont {A.}~\bibnamefont {Tapp}},\ }\bibfield  {title} {\enquote {\bibinfo {title} {Quantum entanglement and the communication complexity of the inner product function},}\ }\href {https://doi.org/10.1016/j.tcs.2012.12.012} {\bibfield  {journal} {\bibinfo  {journal} {Theor. Comput. Sci.}\ }\textbf {\bibinfo {volume} {486}},\ \bibinfo {pages} {11--19} (\bibinfo {year} {2013})}\BibitemShut {NoStop}%
\bibitem [{\citenamefont {Gonzalez-Conde}\ \emph {et~al.}(2024)\citenamefont {Gonzalez-Conde}, \citenamefont {Watts}, \citenamefont {Rodriguez-Grasa},\ and\ \citenamefont {Sanz}}]{GonzalezConde2024}%
  \BibitemOpen
  \bibfield  {author} {\bibinfo {author} {\bibfnamefont {J.}~\bibnamefont {Gonzalez-Conde}}, \bibinfo {author} {\bibfnamefont {T.~W.}\ \bibnamefont {Watts}}, \bibinfo {author} {\bibfnamefont {P.}~\bibnamefont {Rodriguez-Grasa}},\ and\ \bibinfo {author} {\bibfnamefont {M.}~\bibnamefont {Sanz}},\ }\bibfield  {title} {\enquote {\bibinfo {title} {Efficient quantum amplitude encoding of polynomial functions},}\ }\href {https://doi.org/10.22331/q-2024-03-21-1297} {\bibfield  {journal} {\bibinfo  {journal} {Quantum}\ }\textbf {\bibinfo {volume} {8}},\ \bibinfo {pages} {1297} (\bibinfo {year} {2024})}\BibitemShut {NoStop}%
\bibitem [{\citenamefont {{Dom{\'\i}nguez Alonso}}(2014)}]{DominguezAlonso2014}%
  \BibitemOpen
  \bibfield  {author} {\bibinfo {author} {\bibfnamefont {J.~M.}\ \bibnamefont {{Dom{\'\i}nguez Alonso}}},\ }\emph {\bibinfo {title} {{DualSPHysics}: towards high performance computing using {SPH} technique}},\ \href@noop {} {Ph.D. thesis},\ \bibinfo  {school} {Universidad de Vigo}, \bibinfo {address} {Galicia, Spain} (\bibinfo {year} {2014})\BibitemShut {NoStop}%
\bibitem [{\citenamefont {Grover}(1996)}]{Grover1996}%
  \BibitemOpen
  \bibfield  {author} {\bibinfo {author} {\bibfnamefont {L.~K.}\ \bibnamefont {Grover}},\ }\bibfield  {title} {\enquote {\bibinfo {title} {A fast quantum mechanical algorithm for database search},}\ }in\ \href {https://doi.org/10.1145/237814.237866} {\emph {\bibinfo {booktitle} {Proc. 28th ACM SOTC}}}\ (\bibinfo  {publisher} {ACM},\ \bibinfo {address} {New York, NY},\ \bibinfo {year} {1996})\ p.\ \bibinfo {pages} {212–219}\BibitemShut {NoStop}%
\bibitem [{\citenamefont {Wadhia}, \citenamefont {Chancellor},\ and\ \citenamefont {Kendon}(2024)}]{Wadhia2024}%
  \BibitemOpen
  \bibfield  {author} {\bibinfo {author} {\bibfnamefont {V.}~\bibnamefont {Wadhia}}, \bibinfo {author} {\bibfnamefont {N.}~\bibnamefont {Chancellor}},\ and\ \bibinfo {author} {\bibfnamefont {V.}~\bibnamefont {Kendon}},\ }\bibfield  {title} {\enquote {\bibinfo {title} {Cycle discrete-time quantum walks on a noisy quantum computer},}\ }\href {https://doi.org/10.1140/epjd/s10053-023-00795-2} {\bibfield  {journal} {\bibinfo  {journal} {Eur. Phys. J. D}\ }\textbf {\bibinfo {volume} {78}},\ \bibinfo {pages} {29} (\bibinfo {year} {2024})}\BibitemShut {NoStop}%
\bibitem [{\citenamefont {Roffe}(2019)}]{Roffe2019}%
  \BibitemOpen
  \bibfield  {author} {\bibinfo {author} {\bibfnamefont {J.}~\bibnamefont {Roffe}},\ }\bibfield  {title} {\enquote {\bibinfo {title} {Quantum error correction: an introductory guide},}\ }\href {https://doi.org/10.1080/00107514.2019.1667078} {\bibfield  {journal} {\bibinfo  {journal} {Contemp. Phys.}\ }\textbf {\bibinfo {volume} {60}},\ \bibinfo {pages} {226--245} (\bibinfo {year} {2019})}\BibitemShut {NoStop}%
\bibitem [{\citenamefont {Terhal}(2015)}]{Terhal2015}%
  \BibitemOpen
  \bibfield  {author} {\bibinfo {author} {\bibfnamefont {B.}~\bibnamefont {Terhal}},\ }\bibfield  {title} {\enquote {\bibinfo {title} {Quantum error correction for quantum memories},}\ }\href {https://doi.org/10.1103/RevModPhys.87.307} {\bibfield  {journal} {\bibinfo  {journal} {Rev. Mod. Phys.}\ }\textbf {\bibinfo {volume} {87}},\ \bibinfo {pages} {307} (\bibinfo {year} {2015})}\BibitemShut {NoStop}%
\bibitem [{\citenamefont {Cai}\ \emph {et~al.}(2023)\citenamefont {Cai}, \citenamefont {Babbush}, \citenamefont {Benjamin}, \citenamefont {Endo}, \citenamefont {Huggins}, \citenamefont {Li}, \citenamefont {McClean},\ and\ \citenamefont {O'Brien}}]{Cai2023}%
  \BibitemOpen
  \bibfield  {author} {\bibinfo {author} {\bibfnamefont {Z.}~\bibnamefont {Cai}}, \bibinfo {author} {\bibfnamefont {R.}~\bibnamefont {Babbush}}, \bibinfo {author} {\bibfnamefont {S.~C.}\ \bibnamefont {Benjamin}}, \bibinfo {author} {\bibfnamefont {S.}~\bibnamefont {Endo}}, \bibinfo {author} {\bibfnamefont {W.~J.}\ \bibnamefont {Huggins}}, \bibinfo {author} {\bibfnamefont {Y.}~\bibnamefont {Li}}, \bibinfo {author} {\bibfnamefont {J.~R.}\ \bibnamefont {McClean}},\ and\ \bibinfo {author} {\bibfnamefont {T.~E.}\ \bibnamefont {O'Brien}},\ }\bibfield  {title} {\enquote {\bibinfo {title} {Quantum error mitigation},}\ }\href {https://doi.org/10.1103/RevModPhys.95.045005} {\bibfield  {journal} {\bibinfo  {journal} {Rev. Mod. Phys.}\ }\textbf {\bibinfo {volume} {95}},\ \bibinfo {pages} {045005} (\bibinfo {year} {2023})}\BibitemShut {NoStop}%
\bibitem [{\citenamefont {Hayes}\ \emph {et~al.}(2023)\citenamefont {Hayes}, \citenamefont {Croke}, \citenamefont {Messenger},\ and\ \citenamefont {Speirits}}]{Hayes2023}%
  \BibitemOpen
  \bibfield  {author} {\bibinfo {author} {\bibfnamefont {F.}~\bibnamefont {Hayes}}, \bibinfo {author} {\bibfnamefont {S.}~\bibnamefont {Croke}}, \bibinfo {author} {\bibfnamefont {C.}~\bibnamefont {Messenger}},\ and\ \bibinfo {author} {\bibfnamefont {F.}~\bibnamefont {Speirits}},\ }\href@noop {} {\enquote {\bibinfo {title} {Quantum state preparation of gravitational waves},}\ } (\bibinfo {year} {2023}),\ \bibinfo {note} {\href{https://arxiv.org/abs/2306.11073}{arXiv:2306.11073}}\BibitemShut {NoStop}%
\bibitem [{\citenamefont {{S. Croke et al.}}(2025)}]{Croke2025}%
  \BibitemOpen
  \bibfield  {author} {\bibinfo {author} {\bibnamefont {{S. Croke et al.}}},\ }\href@noop {} {} (\bibinfo {year} {2025}),\ \bibinfo {note} {in preparation}\BibitemShut {NoStop}%
\end{thebibliography}
%aipnum4-2.bst 2019-01-14 (MD) hand-edited version of apsrev4-1.bst
\providecommand{\noopsort}[1]{}\providecommand{\singleletter}[1]{#1}%

\end{document}